\documentclass{bmcart}

\usepackage{amsthm,amsmath}
\usepackage{graphicx}
\RequirePackage{natbib}

\usepackage[utf8]{inputenc} 

\startlocaldefs
\endlocaldefs

\begin{document}

\begin{frontmatter}

\begin{fmbox}
\dochead{Research}

\title{Convergence of Economic Growth and the Great Recession as Seen From a Celestial Observatory}

\author[
   addressref={aff1},                   
   corref={aff1},                       
   email={eduede@uchicago.edu}   
]{\inits{ED}\fnm{Eamon} \snm{Duede}}
\author[
   addressref={aff1}
]{\inits{VZ}\fnm{Victor} \snm{Zhorin}}

\address[id=aff1]{
  \orgname{Computation Institute, University of Chicago}, 
  \street{5735 South Ellis Avenue},                     %
  \postcode{60637}                                
  \city{Chicago},                              
  \cny{USA}                                    
}

\end{fmbox}

\begin{abstractbox}

\begin{abstract} 
Macroeconomic theories of growth and wealth distribution have an outsized influence on national and international social and economic policies. Yet, due to a relative lack of reliable, system wide data, many such theories remain, at best, unvalidated and, at worst, misleading. In this paper, we introduce a novel economic observatory and framework enabling high resolution comparisons and assessments of the distributional impact of economic development through the remote sensing of planet earth's surface. Striking visual and empirical validation is observed for a broad, global macroeconomic $\sigma$-convergence in the period immediately following the end of the Cold War. What is more, we observe strong empirical evidence that the mechanisms driving $\sigma$-convergence failed immediately after the financial crisis and the start of the Great Recession. Nevertheless, analysis of both cross-country and cross-state samples indicates that, globally, disproportionately high growth levels and excessively high decay levels have become rarer over time. We also see that urban areas, especially concentrated within short distances of major capital cities were more likely than rural or suburban areas to see relatively high growth in the aftermath of the financial crisis. Observed changes in growth polarity can be attributed plausibly to post-crisis government intervention and subsidy policies introduced around the world. Overall, the data and techniques we present here make economic evidence for the rise of China, the decline of U.S. manufacturing, the euro crisis, the Arab Spring, and various, recent, Middle East conflicts visually evident for the first time.
\end{abstract}


\end{abstractbox}
%

\end{frontmatter}

\section{Introduction \label{sec:intro}}

The rise of more creative and powerful simulation, modeling, and computation along with a superlinear expansion in both the variety and size of data is transforming science, \cite{evans2010philosophy,hey2009fourth}. From high energy physics to cosmology, biology, and genetics, sophisticated instrumentation, massive, high throughput experiments and observatories are increasingly leveraged by scientists and scholars to empirically validate deep, longstanding theories. \cite{aad2012observation,abbott2016observation,covert2004integrating}.

Nevertheless, in many fields, it is still too often the case that deep theory remains untested due to the relative dearth of available, system wide data that would be necessary and sufficient for validation. Data of the commensurate size and shape needed to validate a big theory is often either merely unattainable, too expensive to derive, too elusive to observe, and, for some theories, it may simply be unclear what would so much as count as appropriate data in the first place. Where the requisite data needed to validate a given theory or even observe a theorized phenomenon is lacking, researchers often turn to various methods of indirect detection. An obvious example here is the search for empirical validation of the existence of weakly interacting massive particles \cite{conrad2014indirect,beltran2009deducing}. 

Perhaps not as obviously, many macroeconomic theories, the validity of which carry huge practical, political, and social ramifications, also rely on indirect detection for validation. Data for inferring the state or dynamics of many subnational, national, and global economy-wide phenomena (e.g., employment, income, production, migration, etc.) are gathered through national income and product accounts surveys. Moreover, given that these data are collected almost exclusively through surveying, they are often very small  and of uncertain quality relative to the broad importance of the object of study. For instance, in the United States, data on national unemployment is gathered monthly by the Bureau of Labor and Statistics by surveying roughly 0.02\% of U.S. households. These data are then modified with additional data weighting and statistical adjustments to enhance their stability over time \cite{united2006current}. As a result, these data and the dynamics that are subsequently inferred from them are often unreliable. To give an example, \cite{kaplan2012interstate} demonstrated that much of the observed decrease in US interstate migration was, in fact, a statistical artifact attributable to the Census Bureau's introduction of a seemingly minor change to its procedures for imputing missing data.

Additionally, many nations either lack the necessary organizational or administrative infrastructure to construct accurate and reliable, if nevertheless basic, national accounts or they seek to frustrate the transparency of accounts data for political reasons or economic gain as noted in  \cite{nordhaus2006geography, sinton2001accuracy}.  As a result, many theories that are central to macroeconomics often have an outsized influence on policy even when empirical validation of those theories is missing or, worse, misleading. Extensive discussion of related issues in \cite{rogoff2010growth, herndon2014does, Leamer2007} stresses the importance of rigorous model verification using a wide variety of methods.

While it is not at all clear what the direct observation of, for instance, gross domestic product (GDP), would be, it has been demonstrated that reliable proxies other than variables in national accounts data can been derived through the passive remote sensing of the earth's surface from space. In particular, nighttime luminosity data has been shown convincingly to be a useful and reliable proxy for socioeconomic statistics \cite{ghosh2010shedding,levin2012high,henderson2012measuring,chen2011using,elvidge2012night,pinkovskiy2016lights}. For instance, the strong correlation between aggregate real GDP growth and aggregate changes in luminosity levels was found in \cite{henderson2012measuring} to be highly significant for the period 1992-2008. What is more, \cite{pinkovskiy2016lights} found that variations in GDP explain roughly 75 percent of observed variation in the aggregate nighttime light emissions.

Crucially, recent work has shown robust correlations between the relative intensity of spatially disaggregated nighttime luminosity and GDP at both national and subnational resolutions. Moreover, this work has demonstrated that using nighttime luminosity does particularly well at resolving national and subnational GDP in countries that otherwise lack the administrative statistical infrastructure necessary to derive high-quality national accounts data \cite{nordhaus2015sharper}. Nevertheless, by merely observing statistical correlations between terrestrial light spillage and GDP, the extant literature on leveraging nighttime luminosity data for economic analysis has done little to inform or validate theory about global growth and production dynamics.

In this paper, we argue and demonstrate that passive remote sensing of the earth's surface can be leveraged to do more than merely proxy static accounting. Remote sensing can be used to add robustness to and support the empirical validation of economic theories. As stated above, the macroeconomic tradition for the observation and measurement of phenomena that are predicted by theory has been the employment of surveying and accounting to gather data to which a model can be fit. Apart from precise geographic and temporal resolution, the obvious benefits of these means of data acquisition are the uniform units of measure which are central to the internal structure of a given macroeconomic theory. However, to say nothing of the severe and highly problematic limitations introduced by sample size, the obvious drawbacks surround precision and accuracy of the measured units so that efforts to improve models and, therefore, theory require, at bottom, efforts to improve the institutional infrastructure needed for such measurement. At a deeper level, and for the purposes of validation, the constraint placed on the primary models of these theories by their parameterization and the near axiomatic requirement of specific units (e.g., currency values) of measure in those parameters will ceaselessly frustrate attempts to resolve inconsistencies between theoretical assumptions and observed data. This may seem trivially obvious, but resolving such inconsistencies without substantial modifications to either method or theory (or both) is anything but. These problems are further compounded when the phenomena predicted by theory are, seemingly, only observable by such indirect means.

In what follows, we introduce a novel method with its own, independent set of micro-state assumptions and conditions to detect, sense, or `observe' the theorized macroeconomic phenomena known as `economic convergence' and `divergence'. In particular, we use the highly calibrated, high resolution (pixel-level) heterogeneous magnitude of changes in detectable light spillage over time as a salient proxy for the dynamism of human economic activity. Given that, theories of economic convergence and divergence are, at bottom, concerned with relative changes in economic activity over time, our approach can serve to supply robustness to these theories since the derivation, identification, and measurement of these phenomena can be achieved with novel parameters that are completely independent from those used in all adjacent models. Moreover, in \cite{friedman1992old} it was strongly argued that the real test of economic convergence is a consistent diminution of variance, not among the means of aggregate variables, but among individual enterprises and households, there by, rather indirectly, arguing against the plausibility of observing convergence dynamics in models of aggregate national accounts data.

The remainder of this paper is organized as follows:
Section \ref{sec:setup} describes our data and method (our Celestial Observatory),
Section \ref{sec:results} presents our results,
and Section \ref{sec:conclusion} serves as a summary and discussion.

\section{Setup: data and methods \label{sec:setup}}
There are several flavors of the macroeconomic theory of convergence. In this paper, we are concerned mainly with so called `$\sigma$-convergence' predicting a decrease in the dispersion of income/growth across countries as opposed to the somewhat weaker `$\beta$-convergence' which holds that the economies of poorer countries will grow more rapidly than those of richer ones. In \cite{young2008sigma} it was shown that $\beta$-convergence is a necessary but not sufficient condition for $\sigma$-convergence. There are a number of proposed and plausible explanations for why this dynamic should occur. For instance, in a globalized economy, relatively rich nations will experience faster rates of diminishing returns on freely traded means of production, technologies, and innovations, than poorer countries. As a result, poorer countries will realize faster relative rates of growth than rich countries. In fact, the literature is littered with interpretations of what even counts as growth characteristics for nations \cite{piketty2014capital,durlauf1999new,easterly,nordhaus1996real,rodrik2013unconditional,sala1996classical}. These dynamics are understood to act on very long timescales. For that reason, it is important that growth studies that aim to investigate convergence dynamics work with panel data at the longest timescales available.

Sampling issues notoriously frustrate attempts to observe convergence. For instance, \cite{sala1996classical} observed that the estimated speeds of $\beta$-convergence are so surprisingly similar across data sets, that economists can use a simple rule: economies will converge at a speed of two percent per year. And in \cite{young2008sigma} it is was argued that $\sigma$-convergence did not occur across the United States, or within a majority of the individual U.S. states, from 1970 to 1998. Up to a point, our work, below, seems to affirm the former. However, in the wake of the economic crisis and in the data after 2008, we also observe strong divergence dynamics. So, in certain contexts, at certain statistical moments, and under certain decompositions of the theory itself, our work lends credibility to both. That is, depending on where, when, and how (with what instrument) one looks, one can obverse strong convergence or strong divergence dynamics. Ultimately, we believe that this supports the need for a gentle reassessment of the criteria, data, and instruments employed and deployed. While proposed explanations for why the existence of such theorized dynamics should occur may be intellectually or politically satisfying, from the perspective of economic theory, they ultimately do more to direct attention toward data and model selection rather than to establish the validity of the theory itself. We argue that the following discussion contributes a novel model of an additional, independent data set that provides evidence for the existence of convergence and divergence dynamics.

To generate an independent model for observing convergence and divergence dynamics, we turned to the version 4 DMSP-OLS Nighttime Lights Time Series collected by the US Air Force Weather Agency \cite{doll2008ciesin, baugh2010development}. The cloud-free, stable lights composites were processed using the entirety of the available archive data for the 1992 - 2013 period. This is a 21 year dataset that represents one of the longest panels available for growth studies. The analyzed products are 30 arc second grids, spanning -180 to 180 degrees longitude and -65 to 75 degrees latitude whereby the resulting data arrays include $\approx 730MM$ observed pixels with $\approx 20MM$ non-zero (or active) pixels per year. We note that, while the radiance-calibrated NTL images provide better dynamic resolution, these images are only available at random points in time while, for this study, we seek to concentrate on consistent metrics of economic activity over the whole 1992 - 2013 time period.

Previous work \cite{ghosh2010shedding} has established that the version 4 DMSP-OLS Nighttime Lights Time Series dataset has some substantive benefits in estimating quantifiable economic activity. Nevertheless, like most of the standard, national accounting approaches mentioned above, the use of nighttime lights has limitations. The most practical limitation is introduced by the impossibility of establishing a precise mapping from the rate and intensity of nighttime light leakage to the World Bank's standard metrics (e.g. USD) for GDP. Additionally, the annual composites across years that are recorded by distinct satellites cannot be compared directly with each other due to differences in on-board calibration. Finally, the annual variability of cloud dynamics also affects the statistical reliability and estimate precision, particularly in areas for which there are fewer observations as noted in \cite{free2014trends, norris2015empirical}.  

We began our study by replicating much of the prior literature's work. The extant literature is uniformly committed to analyzing aggregate light intensities. In replicating this literature, we immediately observe that the correlation between changes in aggregate light intensity and world economic growth is merely 0.31 for the period of 1993-2013, which is in line with findings in \cite{henderson2012measuring}. Next, we used a similar technique to separately estimate both country-specific and state-specific aggregate growth metrics for the periods of 1993 - 2006 and 2007 - 2013 with year fixed effects controlling for differences in sensor settings across satellites, as well as taking out the effects of changes in worldwide economic conditions as in \cite{henderson2012measuring}. These results are reported in Table \ref{tab:cross}-\ref{tab:cross-states} for cross-sections of countries and individual U.S. states with average annual growth rates $\hat {y}_{93-06}$ for 1993 - 2006 period and $\hat {y}_{07-13}$ for the 2007 - 2013 period. Using a basic, generalized least-squares procedure, we also estimated the standard deviations of growth across years in  $\sigma( y_{93-06})$, $\sigma (y_{07-13})$. From these, it is obvious that, depending strongly on the particular subsample of countries or years included in studies, the point estimates of growth using prior, extant methods based on night lights can be plausible but misleading and results can be spurious. Nevertheless, and quite surprisingly, aggregate growth metrics have been used exclusively in all previous studies incorporating night lights.

To address these issues while simultaneously keeping the data array of observations as intact and free of adjustments as possible, we apply a differencing technique with zero mean (demeaned) difference centering. In what follows, we describe this method in detail and demonstrate the veracity of this approach. While we cannot completely rid our data of spurious local and global variability due to various measurement factors, this method is appealing both for its simplicity and for its clear empirical findings. Subsequent improvement is certainly possible by, for example, combining and aligning with independent data sets on cloud formation and behavioral dynamics with the possibility of near real-time monitoring of object conditions.

For our study, intensities for nighttime lights are represented by the time-indexed array $X_t(i,j)$, where $t=1992,\dots,2013$ and each element contains the intensity of light detected for a pixel with a given longitude $lon(i)$ and latitude $lat(j)$. 
We perform the following differencing for all pixels to obtain:
$$
\Delta X_t(i,j) = X_t(i,j) - X_{t-1}(i,j), \forall t \in (1993,2013)
$$

At each time step, we then center the differences in pixel light intensity by subtracting the globally (or country, or state-specific) averaged value of differences between years.
\begin{center}
\begin{equation*}
\begin{split}
\Delta X_t^\prime(i,j)*\mathbf{1}(\Delta X_t(i,j)\neq 0)= \\
\Delta X_t(i,j)*\mathbf{1}(\Delta X_t(i,j) \neq 0)  -\frac{\int_{\underline{lon}}^{\overline{lon}} \int_{\underline{lat}}^{\overline{lat}}\Delta X_t(u,v) du dv}{\int_{\underline{lon}}^{\overline{lon}} \int_{\underline{lat}}^{\overline{lat}}\mathbf{1}(\Delta X_t(u,v) \neq 0)dudv},
\end{split}
\end{equation*}
\end{center}
where $\mathbf{1}(\Delta X_t(i,j) \neq 0)$ is an indicator function that is zero for all pixels with no change in intensity and one for all pixels that display a change in intensity.

Thus, we filter away both pixels that do not have any signal and pixels that do not show any difference when demeaned between annual time steps. What is more, this also helps us to avoid a known issue with saturation of light intensity due to detector sensitivity. Static, saturated areas (e.g. urban centers) are automatically filtered with a resultant time series providing a dynamic characterization of detected changes. Obviously, it is plausible that meaningful changes in economic activity in these saturated areas are not detected due to filtering and, therefore, remain hidden. Nevertheless, and perhaps surprisingly, we still observe detectable changes in highly urban and well-developed areas with highly variable light spillage contributing to the detection of economic activity patterns that could be further correlated with other known and observed data on social and economic events. For instance, as small capillaries wend fluctuations toward and away form great estuaries in our data, streams of traffic of higher of lower intensity can be, perhaps, detected spilling in and out of urban central districts on their evening commutes, carrying out freight delivery, or migrating.

Finally, we process our demeaned time series using a simple standard three-state Markov-chain growth model. We use a finite set $S=\{-,0,+\}$ of three possible states: a negative change that is larger than the cross-sectional $\sigma_t$ for year $t$, a positive change that is larger than the cross-sectional $\sigma_t$, and a neutral (no change) state that is within $\sigma_t$.

To understand the characteristics of changes in luminal intensity over time, we compute the transition probabilities $P_t(i|j),\, i,j \in S $ for a state $j$ at time $t-1$ to become state $i$ at  time $t$  as stochastic the matrix:
$$
P_t(+|+); P_t(+|0);P_t(+|-); \sum_i P_t(+|i) =1;
$$
$$
P_t(0|+); P_t(0|0); P_t(0|-); \sum_i P_t(0|i) =1;
$$
$$
P_t(-|+); P_t(-|0); P_t(-|-); \sum_i P_t(-|i) =1;
$$
Next, for each stochastic matrix $P_t$ at time $t$ we find the  asymptotic stationary matrix: $$
\lim_{n\to\infty} P_t^n = \overline P_t
$$
and take the diagonal elements of this matrix as $a_t^{++}$, $a_t^{00}$,  $a_t^{--}$, corresponding to `persistent high growth', `persistent neutral growth', and `persistent high decay', used in subsequent discussion. The benefit of using stationary probabilities conditioned on the state of world at time $t$  is that it allows us to abstract away the transitory effects of stochastic shocks, to reveal how permanent effects from shocks like, for instance, financial crises, shift the world into different trajectories of development. This allows us to test whether variance in the cross-sectional distribution of economically active pixels increase over time. We find that the gap  $a_t^{++}-a_t^{--}$ between the probabilities of persistent, high growth and persistent, high decay provides a good metric for particularly negative (if the gap is negative) or positive shocks that perturb countries on their path to convergence. Obtaining those metrics allows the direct tracking of both the `majority club' (as in \cite{young2008sigma}) of locations that `converge' under specific $\sigma$ bounds and the `minority club' that grow or decay more rapidly and sit about the long tails of the distribution.

\section{Results and discussion \label{sec:results}}

In what follows, we present a number of observations that demonstrate global, national, and subnational convergence and divergence in intensities of terrestrial, nighttime luminosity. We argue that these deltas represent distributional impacts of economic activity and, when these distributional impacts conform to phenomena predicted by broad $\sigma$-convergence/-divergence, we further claim that these observations add robustness to that theory. 

\begin{figure}[t]
\begin{centering}
\includegraphics[width=.9\textwidth]{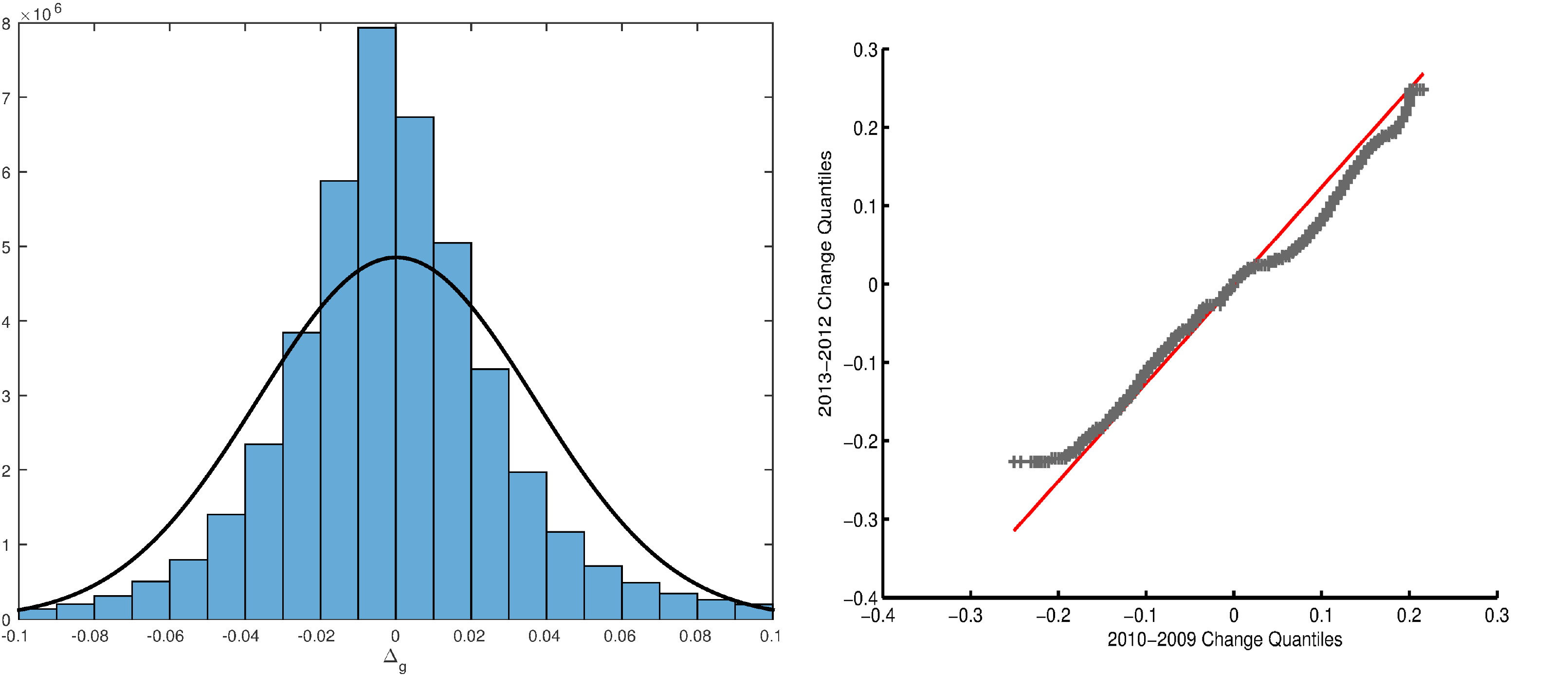}
\end{centering}
      \caption{\csentence{Probability distributions} of observed cumulative zero-mean annual change variations in 1993-2013 period against normal distribution (left) and quantile-quantile-plots across different years (right)}
\label{fig:qqplots}
      \end{figure}

Fig. \ref{fig:qqplots}(left panel) shows a histogram for the demeaned difference distribution representing cumulative change over the 1993 - 2013 period. Our observed change distribution for nighttime light spillage is leptokurtic (excess kurtosis is 8.5) and skewed to the right (where skewness is 0.5 with a standard deviation of 3.68\%). \cite{young2008sigma} studied the U.S. per capita income distribution from 1970 to 1998 and found that excess kurtosis increased from 0.4 to 7.3. This is, indeed, in line with and supported by our findings for night light distribution.  

Fig. \ref{fig:qqplots}(right panel) confirms that samples from different years come from a similar distribution (e.g., AF satellites are looking at the same planet and distribution is stable over time). However, there are important differences across years. For example, pixels in the Middle East that were previously growing in relative intensity suddenly shift into a depressed slide as the region slips into a series of well documented wars and episodes of tremendous civil unrest. 

\begin{figure}[t]
\includegraphics[width=.9\textwidth]{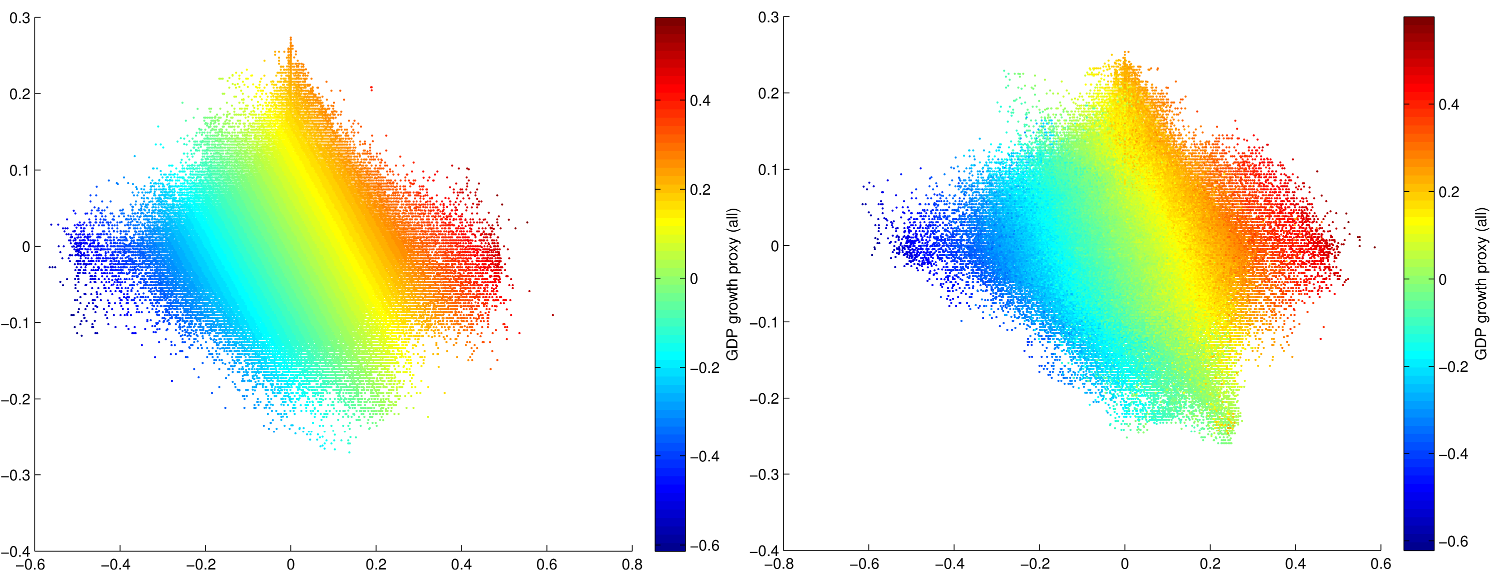}
  \caption{\csentence{Distribution change, effect of Great Recession shock}. \textbf{(Left)}  2003-2006 ($Y$-axis) vs 1992-2002 ($X$-axis),\textbf{(Right)}  2008-2013 ($Y$-axis) vs 1992-2007 ($X$-axis). Total growth over 2013-1992 period is color coded.}
  \label{fig:growth}
      \end{figure}

Fig. \ref{fig:growth}(left panel) shows cumulative change in 1992-2002 (Y axis) vs 2003-2006 (X axis) with color coded pixels based on total intensity growth in 1992-2006 subsample. 
Fig. \ref{fig:growth}(right panel) shows cumulative change in 2008-2013 (Y axis) vs 1992-2007 (X axis) with color coded pixels based on total intensity growth in 1992-2013 sample. A 45-degree (positive sloped) line corresponds to autocorrelation of growth and, therefore, an increase in divergence of growth over time as faster growing pixels persistently outperform slower ones. We see that post-2007 growth shows a noticeable divergence along the systemically high-growth area of the spectrum. However, overall, most points are clustered in either anti-correlated or not correlated quadrants thereby providing support for the idea that general convergence mechanisms are in effect.

\cite{lawrence2013global} found that energy consumption inequality decreased from 0.66 in 1980 to 0.55 in 2010. This finding is in line with the satellite observed production and use of light at night which also indicates similar decreases in inequality after the Cold War and in the run up to the financial crisis of 2008. Unfortunately, comprehensive, post-crisis economic indicators are still lagging and largely unavailable.

We also compute the cumulative \textbf{cross-sectional} standard deviation of our vectorized data array: $$\sigma_t=\sigma(\Delta X_t^\prime(i,j)), \forall t \in (1993,2013)$$.

This approach takes the statistically unreliable, absolute aggregate levels of growth completely out of the picture and centers the focus on the distributional impacts of growth. As a result, $\sigma$-convergence in our framework still captures unconditional estimates of local, high-resolution deviations as in \cite{friedman1992old} while avoiding statistical fallacies required for testing across samples that are grouped by means in a specific year.

\begin{figure}[t]
\includegraphics[width=.9\textwidth]{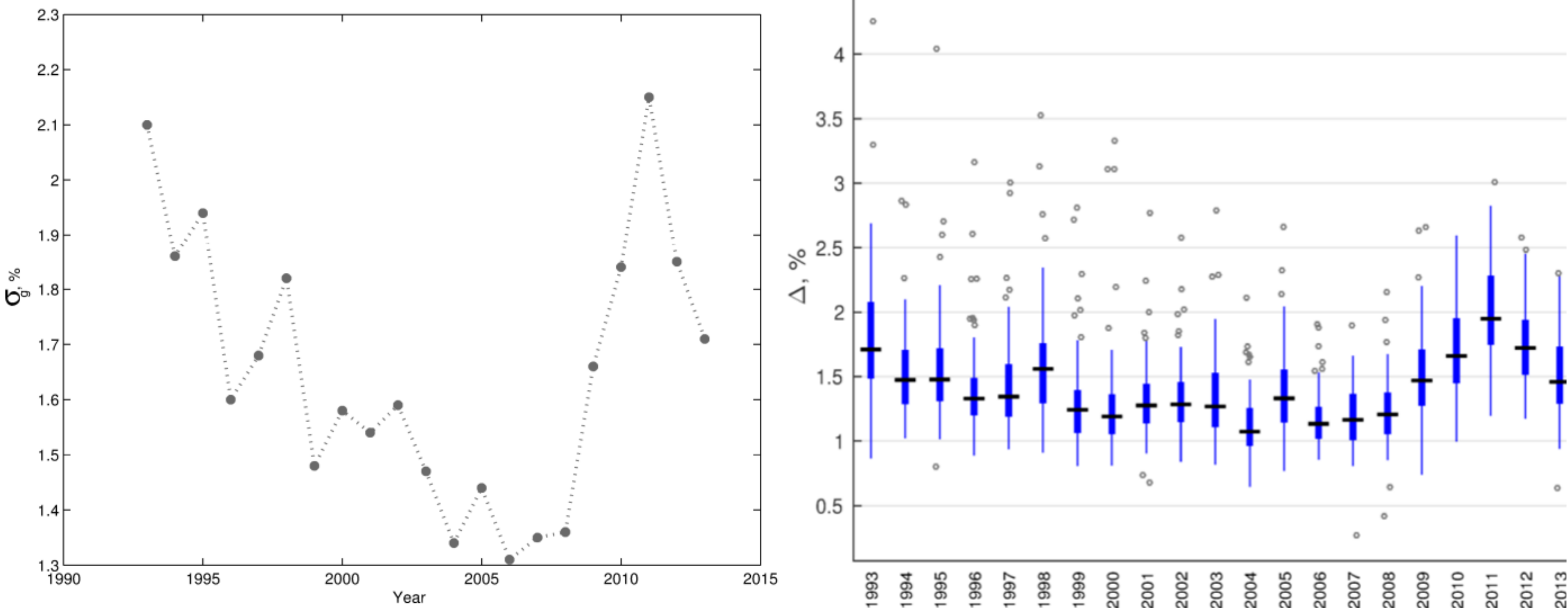}
  \caption{\csentence{Standard deviation  of annual zero-centered change}, \textbf{(Left)} annual global, ($\sigma_g$), percent, \textbf{(Right)} across countries, ($\Delta$) percent  }
  \label{fig:growth_divergence}
      \end{figure}
      
Fig. \ref{fig:growth_divergence} shows global cross-sectional change variability before and after the Great Recession event. We see again, that there is support for a theory of global $\sigma$-convergence kicking in immediately after the end of the Cold War period and before the Great Recession. However, the global cross-sectional provides strong empirical evidence for economic divergence in the years immediately following the economy-wide shocks delivered by the financial crisis with long and persistent recovery dynamics. 

To aid in understanding the sources of these dynamics, we report in Table \ref{tab:cross}-\ref{tab:cross-states} the calculated values for the transition probabilities $a^{++}$, $a^{--}$, $a^{00}$ in our 3-state Markov model. Here we see that both persistent high growth and persistent high decay rates shrank over time, with some prominent exceptions (e.g., Syria). In general, moreover, the gap between high growth and high decay became larger indicating that the mechanisms driving $\sigma$-convergence failed immediately after the financial crisis. This observation highlights another strength of our approach. Namely, it provides indicators for the direction of research attention and aids in the development of hypotheses. For instance, perhaps the observed breakdown in $\sigma$-convergence dynamics can be attributed to intentional suppression of extreme volatility through post-crisis monetary policies. Such policies may be directly implicated in decreasing the probability of failure to a large degree while also decreasing slightly the probability of higher success.

\begin{figure}[h!]
\includegraphics[width=.9\textwidth]{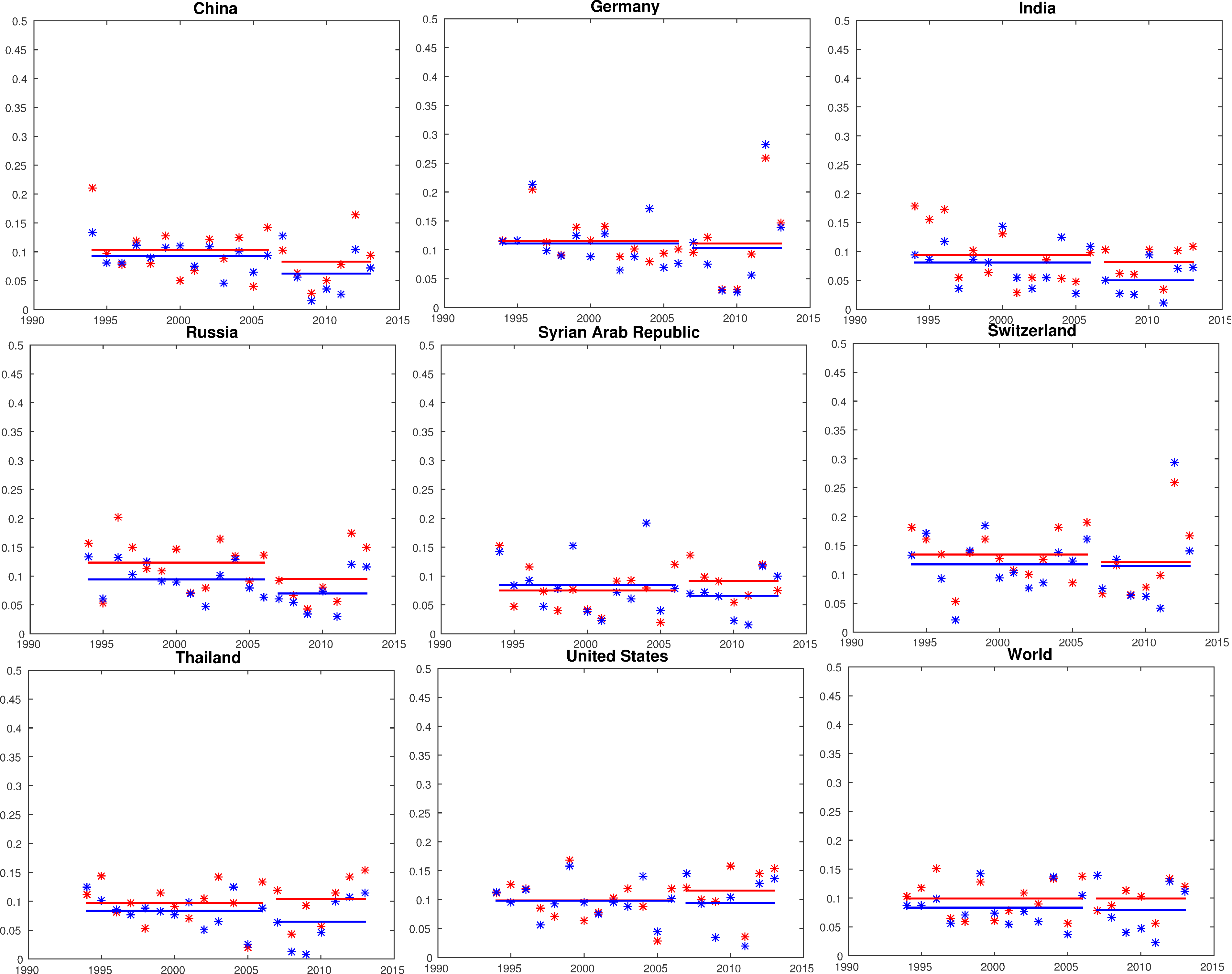}
  \caption{\csentence{Markov transition probabilities across selected countries}, red markers are for $a^{++}$ values, blue markers are for $a^{--}$ values,  solid blue and red lines are mean values over 1993-1006 and 2007-2013 periods, percent  }
  \label{fig:markov-country}
      \end{figure}  
          
\begin{figure}[h!]
\includegraphics[width=.9\textwidth]{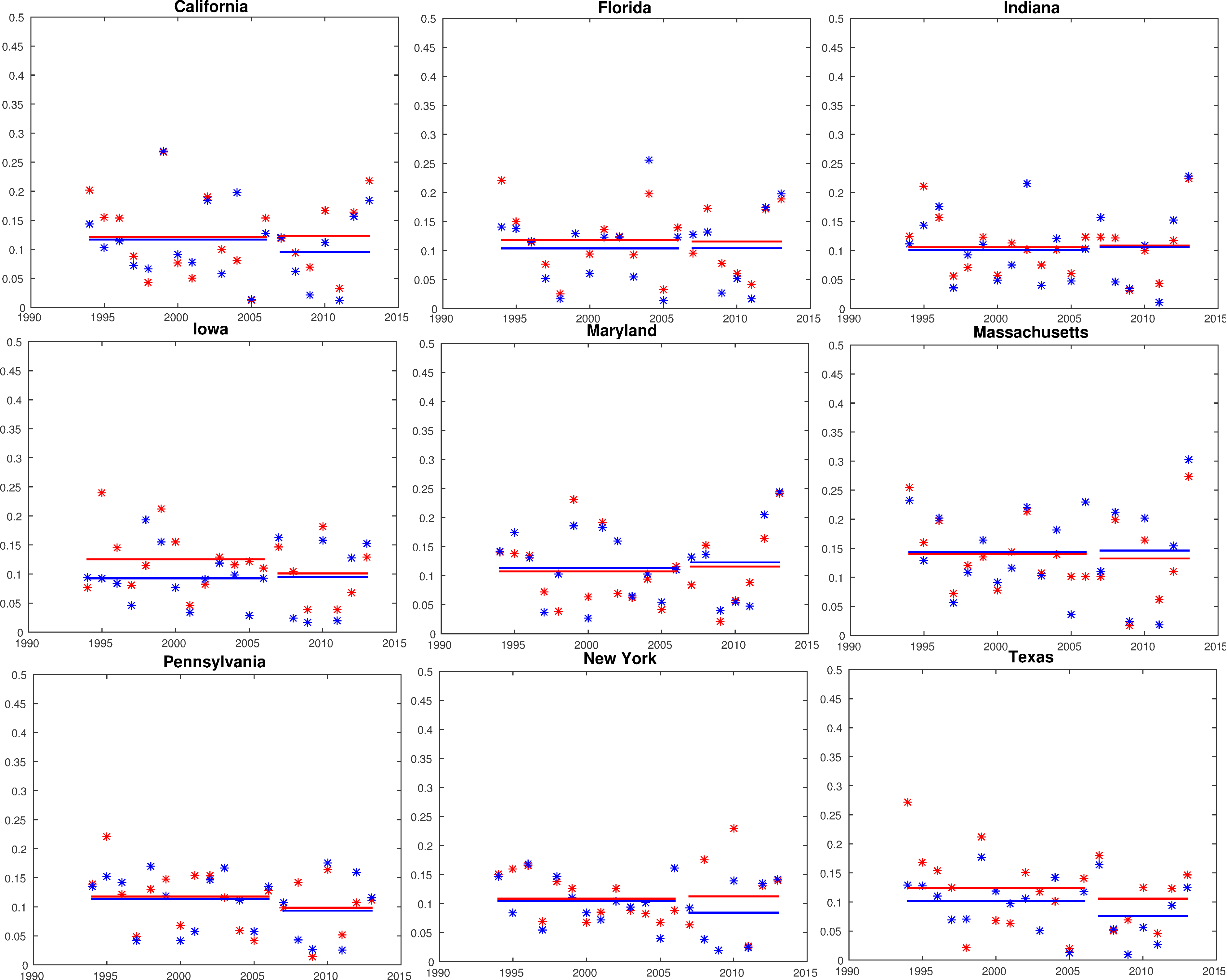}
  \caption{\csentence{Markov transition probabilities across selected US states}, red markers are for $a^{++}$ values, blue markers are for $a^{--}$ values,  solid blue and red lines are mean values over 1993-1006 and 2007-2013 periods, percent  }
  \label{fig:markov-state}
      \end{figure}  
      
Figures \ref{fig:markov-country}-\ref{fig:markov-state} show the annual Markov transition probabilities for a selected subset of countries and U.S. states. Red and blue dots represent the observed transition probabilities for high growth areas to high growth areas and high decay areas to high decay areas respectively. The solid red and blue lines represent the mean probabilities for a given time period with a break at the end of the 1993 - 2006 subsample and the start of the sample for 2007 - 2013. The mean observations tell a fairly consistent story and confirm that there is broad heterogeneity in the observed variation in growth and decay probability outcomes across nations after the global financial crisis. The individual observations also tell a series of interesting stories. For example, Switzerland appears to have been particularly sensitive to the financial crises that occurred in the observed period, including the Asian crisis (1997) and Euro zone crisis (2009 - 2011). Looking at the United States, we observe that the high growth probabilities and high decay probabilities moved in, more or less, lockstep through most of the 1993 - 2006 period but that, after the financial crisis, a gap opened up whereby the high growth probabilities climbed slightly while the high decay probabilities look rather unchanged. One could reasonably hypothesize that this uptick in growth-growth probability is due to massive stimulus programs enacted in direct response to the broad financial meltdown.

\begin{table}[ht]
\centering \tiny
\caption{Cross-country comparison of average aggregate  annual growth and Markov transition probabilities,  percent}\label{tab:cross}
\begin{tabular}{|c|c|c|c|c|c|c|c|c|c|c|}
\hline 

country name & $\hat {y}_{93-06}$ & $\hat {y}_{07-13}$ & $\sigma(y_{93-06})$  & $\sigma(y_{07-13})$ & $a^{++}_{93-06}$ & $a^{++}_{07-13}$ & $a^{--}_{93-06}$ & $a^{--}_{07-13}$& $a^{00}_{93-06}$ & $a^{00}_{07-13}$\\

\hline Afghanistan&6.07&6.31&20.1&12.4&10.4&11.1&7.6&6.9&82.0&82.0\\
\hline Albania&13.98&12.88&29.3&25.2&8.9&12.1&7.4&10.8&83.7&77.1\\
\hline Algeria&1.93&3.00&1.7&1.0&10.2&10.1&8.1&7.8&81.7&82.1\\
\hline Armenia&5.43&5.58&18.5&12.8&11.3&10.6&9.8&7.1&78.9&82.3\\
\hline Austria&2.60&3.37&4.8&3.3&11.9&9.8&10.3&8.8&77.8&81.4\\
\hline Azerbaijan&1.43&3.37&10.3&6.1&11.8&7.7&9.7&5.3&78.5&87.0\\
\hline Bangladesh&1.88&3.08&3.2&2.5&9.3&10.1&7.6&4.5&83.1&85.4\\
\hline Belgium&1.80&2.91&1.8&1.2&14.3&14.6&14.2&12.7&71.5&72.7\\
\hline Brazil&2.60&3.44&3.2&2.1&11.7&12.2&10.5&9.3&77.9&78.5\\
\hline Bulgaria&3.55&3.89&9.3&6.5&11.9&9.5&10.2&6.5&77.9&84.1\\
\hline Cambodia&6.87&11.97&13.3&31.2&9.7&9.5&5.9&2.4&84.4&88.1\\
\hline China&2.07&3.11&1.0&0.8&10.4&8.3&9.3&6.2&80.4&85.5\\
\hline Colombia&2.09&3.13&2.6&1.7&9.7&11.6&10.2&7.5&80.1&81.0\\
\hline Costa Rica&4.17&4.79&6.1&5.1&7.7&9.2&10.2&6.7&82.1&84.1\\
\hline Croatia&6.84&5.16&10.6&6.5&11.1&11.3&10.9&9.4&78.0&79.4\\
\hline Cuba&4.60&4.86&8.9&6.6&9.0&10.9&8.7&7.2&82.3&82.0\\
\hline Czech Republic&2.48&3.51&9.9&7.1&13.0&10.3&12.4&10.7&74.5&79.0\\
\hline Ecuador&2.38&3.60&2.6&2.3&10.6&9.2&9.4&8.0&80.1&82.8\\
\hline Egypt&2.09&3.17&0.9&1.0&9.3&11.9&10.7&8.2&79.9&79.9\\
\hline Eritrea&4.77&4.37&6.4&5.3&10.2&11.9&10.7&6.2&79.1&81.9\\
\hline Estonia&9.82&6.56&17.6&10.2&12.4&9.2&9.5&6.5&78.1&84.3\\
\hline Ethiopia&6.06&5.89&9.3&7.3&10.8&13.2&9.5&6.1&79.7&80.7\\
\hline Finland&4.35&4.50&10.3&6.4&12.1&10.9&10.5&10.5&77.4&78.6\\
\hline France&1.76&2.87&1.0&0.7&12.4&11.1&9.2&9.6&78.5&79.3\\
\hline Georgia&3.72&4.92&15.4&11.6&12.5&13.1&10.1&7.6&77.4&79.4\\
\hline Germany&1.62&2.92&1.6&1.6&11.5&11.1&11.1&10.3&77.4&78.6\\
\hline Greece&2.50&3.24&3.9&2.3&8.8&11.4&9.5&8.5&81.7&80.1\\
\hline Guatemala&2.86&3.71&2.5&2.2&8.6&8.6&9.3&5.2&82.2&86.2\\
\hline Hong Kong&1.20&2.51&3.2&3.2&12.8&13.6&13.7&13.6&73.5&72.8\\
\hline Hungary&5.24&4.24&12.6&7.3&12.3&9.2&11.9&6.4&75.8&84.3\\
\hline India&1.77&2.93&0.5&0.6&9.4&8.2&8.1&5.0&82.5&86.8\\
\hline Indonesia&3.98&4.70&8.1&6.1&8.5&9.3&8.1&6.2&83.4&84.5\\
\hline Iran &1.91&2.93&0.8&0.7&10.8&10.4&10.0&7.7&79.2&81.9\\
\hline Iraq&2.17&3.83&2.9&2.9&9.7&9.9&9.0&6.5&81.3&83.6\\
\hline Ireland&3.00&3.43&5.1&3.0&11.3&11.8&10.7&8.2&77.9&80.1\\
\hline Israel&1.84&2.97&1.0&0.8&13.6&13.3&13.5&11.0&72.9&75.7\\
\hline Italy&1.86&2.93&1.1&0.8&13.1&12.9&11.2&10.9&75.8&76.1\\
\hline Japan&1.68&2.83&0.5&0.5&12.7&11.5&12.2&10.0&75.2&78.5\\
\hline Jordan&2.20&3.21&2.2&1.4&10.5&11.2&9.8&8.2&79.7&80.7\\
\hline Kazakhstan&2.31&3.47&5.0&3.4&13.9&10.4&8.6&7.3&77.5&82.3\\
\hline Korea&2.03&3.03&1.7&1.0&10.5&11.5&9.5&10.0&80.1&78.5\\
\hline Kuwait&3.18&3.94&7.8&4.6&7.8&11.0&7.9&9.9&84.3&79.1\\
\hline Kyrgyzstan&3.28&4.04&9.5&5.7&12.6&9.8&8.3&8.9&79.1&81.2\\
\hline Latvia&8.36&5.33&14.2&9.2&12.2&10.2&11.3&8.3&76.5&81.5\\
\hline Libya&1.86&2.92&2.6&1.5&10.1&9.8&10.0&8.8&79.9&81.5\\
\hline Lithuania&7.40&5.51&16.4&12.6&12.2&10.9&9.8&6.3&78.0&82.8\\
\hline Luxembourg&1.80&3.12&6.4&3.5&12.5&13.3&13.5&12.7&74.1&74.1\\
\hline Mexico&1.87&2.95&1.0&0.9&9.5&9.7&9.2&7.3&81.3&83.0\\
\hline Mongolia&3.08&4.08&6.5&4.0&11.6&10.6&8.3&8.0&80.1&81.4\\
\hline Morocco&2.07&3.06&1.2&0.8&10.4&11.2&10.6&8.8&79.1&80.0\\
\hline Nepal&2.26&3.62&3.1&4.9&10.2&11.0&10.1&7.3&79.8&81.6\\
\hline Netherlands&1.70&2.85&0.9&0.8&14.0&14.5&13.6&12.6&72.4&72.9\\
\hline Nicaragua&3.24&4.39&4.8&4.6&10.2&9.9&10.0&6.4&79.7&83.7\\
\hline Norway&4.85&4.71&11.6&6.8&11.5&12.4&9.0&10.0&79.5&77.6\\
\hline Oman&3.10&3.54&3.2&1.8&10.7&8.6&9.7&6.7&79.6&84.6\\
\hline Pakistan&1.48&2.74&2.2&2.4&8.9&8.1&8.1&7.8&83.0&84.1\\
\hline Palestine&1.80&3.02&1.7&1.2&11.6&13.5&11.1&11.4&77.2&75.1\\
\hline Poland&5.09&4.61&13.0&8.6&12.4&10.2&12.6&9.6&75.1&80.2\\
\hline Portugal&2.21&3.14&1.5&1.3&10.3&10.7&8.8&9.4&80.9&79.8\\
\hline Puerto Rico&1.53&2.73&1.8&1.4&14.6&14.3&15.5&11.5&70.0&74.2\\
\hline Qatar&2.64&3.84&3.9&3.2&10.3&10.5&11.1&6.7&78.6&82.8\\
\hline Romania&3.88&4.57&6.3&6.3&11.9&9.8&9.3&7.5&78.8&82.7\\
\hline Russia&1.95&3.06&2.5&1.5&12.3&9.5&9.4&7.0&78.2&83.5\\
\hline Saudi Arabia&1.91&2.97&0.6&0.5&10.1&10.4&9.7&6.9&80.3&82.8\\
\hline Serbia&2.95&3.51&5.5&3.6&11.6&12.1&8.9&9.0&79.5&78.9\\
\hline Singapore&1.66&2.81&0.7&0.5&NaN&11.2&NaN&11.6&NaN&77.3\\
\hline Slovakia&2.28&3.38&8.3&4.4&12.8&10.4&11.3&7.8&75.9&81.7\\
\hline Slovenia&4.62&4.11&8.7&5.3&10.4&10.1&9.3&9.4&80.3&80.5\\
\hline Spain&1.81&2.91&0.7&0.8&11.5&11.9&10.8&10.6&77.7&77.5\\
\hline Sudan&2.38&3.17&1.3&0.9&9.8&11.8&8.9&6.8&81.3&81.4\\
\hline Sweden&2.80&3.44&6.8&3.8&12.7&10.9&10.6&10.0&76.7&79.1\\
\hline Switzerland&2.43&3.33&7.8&4.3&13.4&12.1&11.7&11.5&74.8&76.4\\
\hline Syria&1.86&2.65&3.1&4.1&7.5&9.2&8.4&6.6&84.0&84.2\\
\hline Taiwan&2.24&3.16&3.2&2.0&12.8&13.9&14.6&9.7&72.6&76.4\\
\hline Tajikistan&0.41&2.88&5.5&5.0&13.1&8.5&10.1&9.2&76.8&82.3\\
\hline Thailand&2.47&3.85&2.3&3.9&9.7&10.3&8.4&6.5&82.0&83.2\\
\hline Tunisia&2.63&3.32&3.4&2.0&7.8&8.7&10.7&6.7&81.5&84.6\\
\hline Turkey&2.93&3.81&4.2&3.4&10.2&12.9&8.9&7.4&80.9&79.7\\
\hline Turkmenistan&2.15&3.17&3.7&2.4&10.7&9.4&6.6&6.2&82.7&84.4\\
\hline Ukraine&0.57&2.94&7.9&5.2&13.8&9.3&10.2&5.1&76.0&85.6\\
\hline United Arab Emirates&2.13&3.11&1.1&0.8&9.0&10.9&9.1&8.4&81.9&80.7\\
\hline United Kingdom&1.46&2.79&1.6&1.1&11.5&11.7&11.5&11.2&77.0&77.1\\
\hline United States&1.73&2.86&0.6&0.6&9.8&11.6&9.8&9.4&80.4&79.0\\
\hline Uzbekistan&1.19&2.72&2.7&1.6&11.7&7.9&8.9&7.6&79.4&84.5\\
\hline Venezuela&1.97&3.29&1.7&2.0&10.3&10.4&10.3&8.0&79.4&81.7\\
\hline Vietnam&4.43&4.14&4.8&2.8&8.3&9.6&8.7&7.0&83.1&83.4\\
\hline Western Sahara&3.00&3.50&2.2&1.6&11.1&10.6&10.8&8.3&78.2&81.1\\
\hline World&1.66&2.78&0.0&0.0&9.9&9.9&8.3&7.9&81.8&82.2\\
\hline Yemen&2.88&3.48&2.5&2.2&7.6&10.5&7.2&6.7&85.2&82.8\\
\hline 
\end{tabular} 
\end{table} 

\begin{table}[ht]
\centering \tiny
\caption{Cross-state comparison of average aggegate annual growth and Markov transition probabilities,  percent}\label{tab:cross-states}
\begin{tabular}{|c|c|c|c|c|c|c|c|c|c|c|}
\hline 

country name & $\hat {y}_{93-06}$ & $\hat {y}_{07-13}$ & $\sigma(y_{93-06})$  & $\sigma(y_{07-13})$ & $a^{++}_{93-06}$ & $a^{++}_{07-13}$ & $a^{--}_{93-06}$ & $a^{--}_{07-13}$& $a^{00}_{93-06}$ & $a^{00}_{07-13}$\\
\hline Alabama&1.71&2.86&1.3&0.9&10.9&10.3&10.6&8.9&78.5&80.7\\
\hline Arizona&1.75&2.87&0.7&0.6&10.6&14.0&10.9&10.7&78.5&75.3\\
\hline Arkansas&1.96&2.97&2.0&1.3&10.4&11.0&10.2&7.4&79.3&81.5\\
\hline California&1.55&2.76&0.9&0.7&12.1&12.3&11.7&9.5&76.2&78.1\\
\hline Colorado&1.56&2.74&3.1&3.5&13.5&12.6&10.9&10.4&75.6&77.0\\
\hline Connecticut&1.75&3.11&2.2&3.5&15.0&13.8&13.6&13.5&71.4&72.6\\
\hline Delaware&1.52&2.91&4.7&2.5&11.2&12.5&10.5&14.3&78.3&73.2\\
\hline Florida&1.68&2.81&0.3&0.3&11.8&11.5&10.4&10.4&77.8&78.1\\
\hline Georgia.&1.86&2.91&1.4&0.9&10.5&9.7&10.9&8.6&78.6&81.7\\
\hline Idaho&2.32&3.17&8.4&7.0&11.7&10.2&10.4&7.6&77.9&82.2\\
\hline Illinois&2.08&3.12&2.5&2.3&11.0&10.8&10.7&9.3&78.3&79.9\\
\hline Indiana&2.29&3.26&4.7&3.2&10.6&10.8&10.1&10.5&79.3&78.6\\
\hline Iowa&2.84&3.92&4.2&4.9&12.5&10.1&9.2&9.4&78.2&80.5\\
\hline Kansas&1.69&2.96&3.4&2.8&14.5&11.7&10.2&8.0&75.2&80.4\\
\hline Kentucky&1.51&2.90&4.7&2.4&11.6&9.5&11.0&7.6&77.3&82.9\\
\hline Louisiana&1.68&2.84&0.8&0.6&9.8&8.7&9.8&7.0&80.5&84.3\\
\hline Maine&3.56&4.88&7.2&9.4&10.2&10.6&9.5&7.6&80.3&81.8\\
\hline Maryland&1.32&2.79&3.9&2.2&10.7&11.6&11.3&12.3&78.0&76.2\\
\hline Massachusetts&1.79&3.08&1.0&2.2&14.0&13.2&14.4&14.6&71.6&72.2\\
\hline Michigan&2.31&3.77&2.4&4.5&11.2&10.8&10.8&8.5&78.0&80.7\\
\hline Minnesota&2.32&3.73&2.5&4.4&11.7&9.1&9.7&7.4&78.6&83.5\\
\hline Mississippi&1.85&2.92&1.3&1.0&11.0&8.0&10.1&7.6&78.9&84.3\\
\hline Missouri&1.97&3.01&2.9&1.8&11.6&9.9&10.0&7.3&78.4&82.8\\
\hline Montana&4.58&4.87&15.6&12.1&13.7&9.8&9.4&5.8&76.9&84.4\\
\hline Nebraska&2.34&3.50&4.4&5.3&12.2&14.8&11.2&10.9&76.7&74.3\\
\hline Nevada&1.73&2.86&2.2&1.7&12.7&11.7&10.9&9.2&76.5&79.1\\
\hline New Hampshire&1.98&3.50&1.7&4.6&12.1&11.9&11.7&11.4&76.2&76.8\\
\hline New Jersey&1.62&2.87&2.5&1.8&13.4&12.0&13.1&12.4&73.5&75.6\\
\hline New Mexico&1.61&2.84&1.2&1.0&12.7&11.1&10.7&8.7&76.6&80.2\\
\hline New York&2.11&3.48&2.3&4.0&10.9&11.2&10.5&8.4&78.6&80.3\\
\hline North Carolina&1.54&2.83&2.4&1.3&9.9&11.6&11.4&8.2&78.8&80.2\\
\hline North Dakota&4.33&5.60&10.2&8.3&15.5&8.7&9.5&5.3&75.1&86.0\\
\hline Ohio&2.02&3.13&4.0&2.9&11.3&10.3&10.5&9.1&78.1&80.7\\
\hline Oklahoma&1.64&2.85&1.7&1.2&11.3&12.8&10.0&8.1&78.7&79.0\\
\hline Oregon&1.70&2.76&3.4&3.0&12.3&11.8&10.1&11.3&77.6&76.9\\
\hline Pennsylvania&2.13&3.27&3.4&3.3&11.8&9.8&11.3&9.3&76.9&80.8\\
\hline Rhode Island&1.51&2.88&2.2&2.5&13.9&12.2&12.5&14.0&73.6&73.8\\
\hline South Carolina&1.73&2.85&1.0&0.7&10.6&10.4&11.2&8.5&78.2&81.1\\
\hline South Dakota&3.18&4.00&5.8&5.2&13.4&12.3&12.3&9.0&74.2&78.7\\
\hline Tennessee&1.47&2.81&2.9&1.7&11.0&10.1&10.8&7.4&78.1&82.5\\
\hline Texas&1.67&2.87&0.9&0.7&12.4&10.6&10.2&7.5&77.4&81.9\\
\hline Utah&1.57&2.84&5.1&4.0&11.0&10.3&10.5&9.1&78.5&80.6\\
\hline Vermont&2.35&3.82&3.7&7.1&10.2&8.2&9.1&6.6&80.7&85.2\\
\hline Virginia&1.56&2.87&3.8&2.0&10.6&8.2&11.0&7.4&78.4&84.5\\
\hline Washington&1.85&2.76&6.8&5.4&13.6&10.4&10.9&8.2&75.5&81.4\\
\hline West Virginia&1.75&2.93&4.1&2.3&11.2&10.1&10.6&7.3&78.2&82.7\\
\hline Wisconsin&2.20&3.96&2.0&4.9&10.5&10.3&10.0&8.2&79.5&81.5\\
\hline Wyoming&1.63&3.10&6.1&7.4&13.4&13.0&9.4&10.1&77.2&76.9\\
\hline 
\end{tabular} 
\end{table}

Apart from interrogating the calculated transition probabilities, much can be learned simply by looking at individual pixel-level data. So, in order to facilitate merely looking, we produced maps of the global distributional impacts with color-coded values of $\Delta \mathbf{X}(i,j) = \sum_{t=1993}^{t=2013}\Delta X_t^\prime(i,j)$.

\begin{figure}[t]
\includegraphics[width=0.9\textwidth]{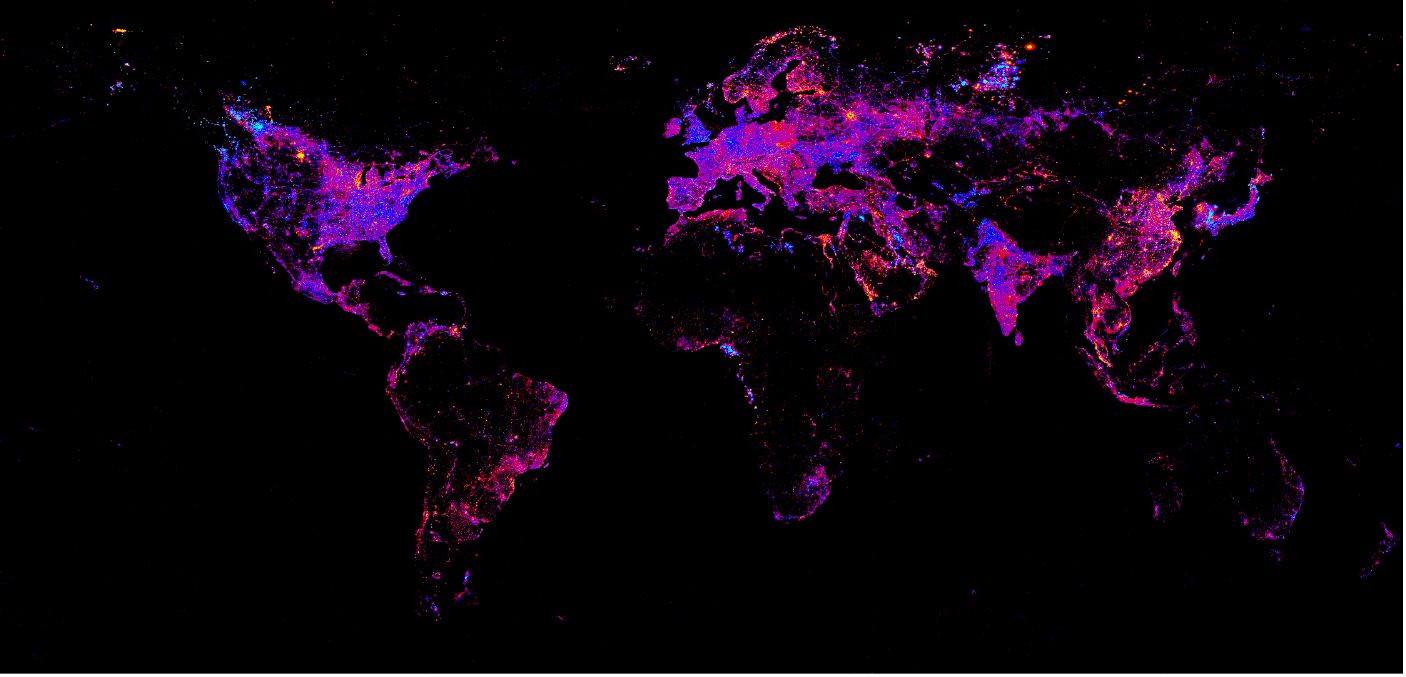}
  \caption{\csentence{Annual average  global mean-centered change, 1993-2013}.  Blue-coded areas correspond to net negative dynamics (decay) and red-colored areas to net positive dynamics (growth). Neutral changes are not distinguishable from the black background. Extreme outliers for negative and positive change are marked with cyan and orange correspondingly }
\label{fig:global}
      \end{figure}

Fig. \ref{fig:global} shows the total, globally mean-centered change over a period of 22 years spanning 1992-2013. Blue-coded areas correspond to net negative dynamics (decay) and red-colored areas to net positive dynamics (growth). Neutral changes are not distinguishable from the black background. It is important stress that coded deviations represent movements away from the global zero-mean centered panel time series. This means that relative decline (blue-colored areas) or relative growth (red-colored areas) do not necessarily translate to absolute decline of economic activity in a given area. This is, of course, consistent with broad mechanics of $\sigma$-convergence.

Looking closely, the major economic transformations following the end of Cold War are clearly visible with, for example, the net positive rise of China and the decline of US industry, capital transfers within the European Union with net beneficiaries in Eastern Europe and Scandinavia, growth in India and China alongside the relative decline in Pakistan and Syria. The development of shale oil drilling along side conventional oil extraction techniques represent important factors in economic production and development and their impact can be clearly observed in the northern United States and Russia.

To demonstrate the resolution of our observatory, we also conduct distributional impact studies with $\sigma$-convergence tests \textbf{within} particular countries and states of the United States across our full panel data time series. For these studies, we zero-mean centered our time series data for each, particular, country or state as a whole for country/state-specific fixed effects.
   \begin{figure}[t]
   \includegraphics[width=.9\textwidth]{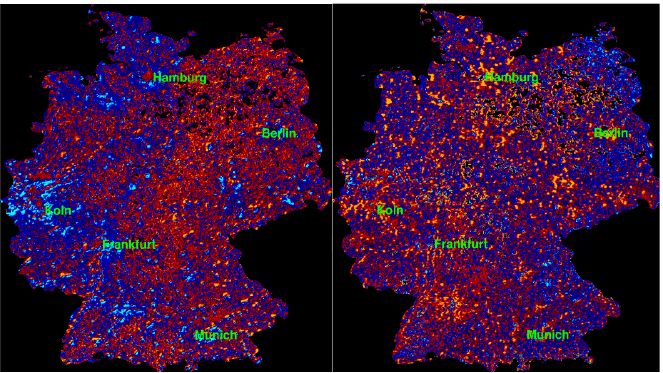}
  \caption{\csentence{Annual average mean-centered change, Germany}. \textbf{(Left)} 1993-2006 period,\textbf{(Right)} 2007-2013  period}
\label{fig:germany}
      \end{figure}
      
Fig. \ref{fig:germany} represents a decomposition of the total mean-centered change into annual averages for the period of 1993 - 2006 (left panel) as compared to 2007 - 2013 (right panel) for Germany. There is a distinct, observable contrast between relatively larger growth in central and Eastern Germany before the global financial crisis and a shift in growth toward historically more developed centers in the western regions following the crisis. 

      \begin{figure}[t]
        \includegraphics[width=.9\textwidth]{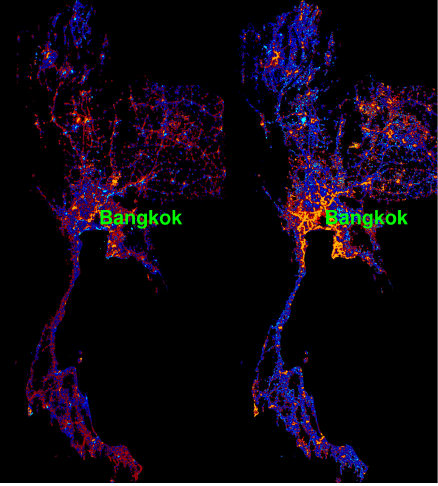}
\caption{\csentence{Annual average mean-centered change, Thailand}. \textbf{(Left)} 1993-2006 period,\textbf{(Right)} 2007-2013  period}
\label{fig:thailand}
      \end{figure}

Fig. \ref{fig:thailand}(left panel, right panel) represents the same period decompositions for Thailand, one of the more successful developing nations with notably sustained high levels of growth. Interestingly, there is an obvious and quite sharp contrast between an observably equatable or diffuse growth distribution before the crisis and a post crisis concentration of growth toward and around central Bangkok.

      \begin{figure}[t]
             \includegraphics[width=.9\textwidth]{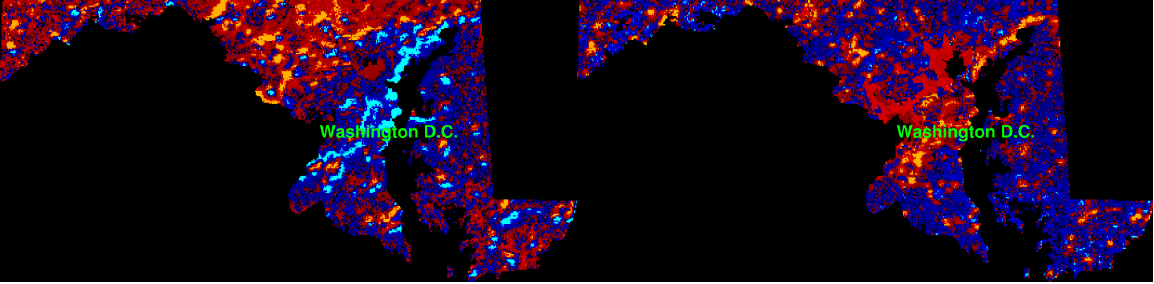}
  \caption{\csentence{Annual average mean-centered change, Maryland}. \textbf{(Top)} 1993-2006 period,\textbf{(Bottom)} 2007-2013  period}
\label{fig:maryland}
      \end{figure}
      
Fig. \ref{fig:maryland}(left, right) represents the same period decomposition for the state of Maryland (U.S.A.) that surrounds the nation's capital, Washington, D.C.. The magnitude of observable urban renewal is striking. We also note a similar phenomenon for the city of Chicago, as well as other major metropolitan areas in the U.S. and around the world. There are, nevertheless, equally striking exceptions.

      \begin{figure}[t]
       \includegraphics[width=.9\textwidth]{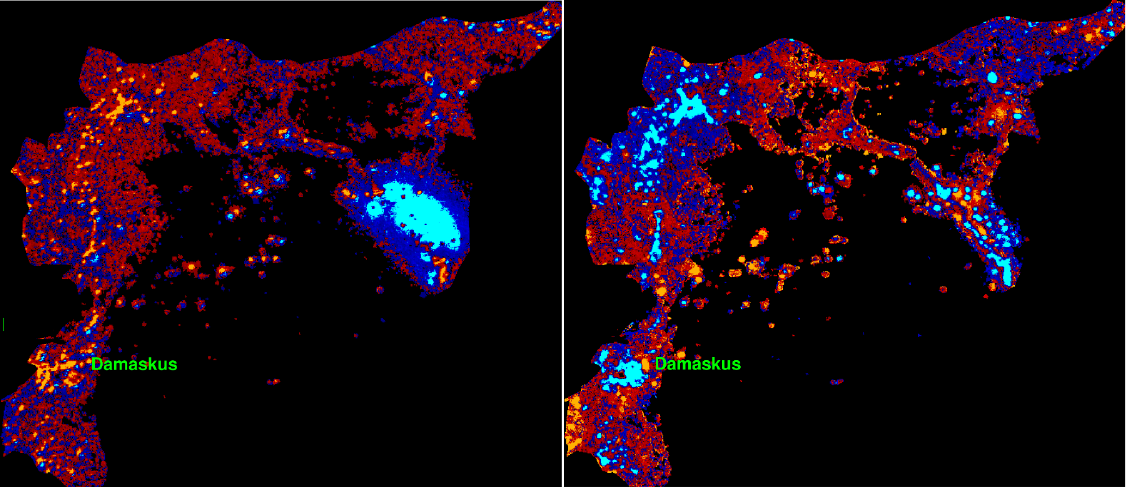}
  \caption{\csentence{Annual average mean-centered change, Syria}. \textbf{(Left)} 1993-2006 period,\textbf{(Right)} 2007-2013  period}
\label{fig:syria}
      \end{figure}
            \begin{figure}[t]
                   \includegraphics[width=.9\textwidth]{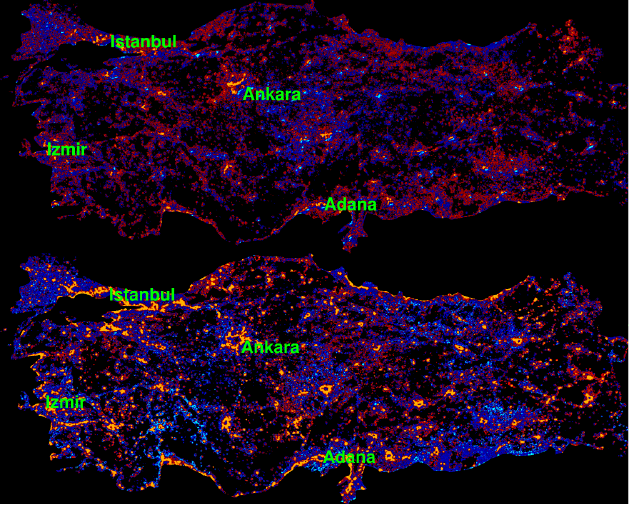}
  \caption{\csentence{Annual average mean-centered change, Turkey}. \textbf{(Top)} 1993-2006 period,\textbf{(Bottom)} 2007-2013  period}
\label{fig:turkey}
      \end{figure}
      
Fig. \ref{fig:syria}(left panel, right panel)-Fig. \ref{fig:turkey}(top panel, bottom panel) represents Syria and Turkey. The change in growth polarity for the once prosperous province of Aleppo, since captured by insurgents, as well as new areas of growth away from the central government are clearly visible and highlight a possible use of this observatory and method for humanitarian purposes. Namely, the early detection of on-the-ground social crises.

Our approach directly exposes a substantial amount of policy relevant heterogeneity in economic development both inside and across countries. Moreover, this heterogeneity is demonstrably impossible to observe using only mean-aggregated data and serves to further emphasize the contributions our observatory and approach can make to theory. While the correlation between photons emitted as a byproduct of economic transactions and economic development has been described in the literature, technology improvements should, in theory, lead to increases in terrestrial light spillage as a result of falling costs. For instance, in \cite{nordhaus1996real}, the amount of labor required to pay for a lumen-hour of light consumption was estimated over time and shown to fall precipitously. However, technologies and energy conservation measures can introduce ambiguous observational effects and further frustrate attempts to make sense of observations. On the one hand, conservation necessarily leads to reductions in light spillage due to, for instance, improved efficiencies in building lighting management (i.e. automatically turning off lights at night). On the other hand, technologies can be implicated in increased luminal production due to more efficient energy-to-light conversion mediums (as, for example, switching from incandescent lighting to LED) thereby driving down the cost (in both dollars and carbon) of leaving the lights on. Of course, increasing urban population densities are a factor. Yet, as we can see from Fig. \ref{fig:thailand}-Fig. \ref{fig:maryland}, there is a visible effect from the growth of suburbia and general urban sprawl, suggesting that urban commute light spillage does, at least partially, capture the effects of urban growth. Nevertheless, applying proper spatial statistical methods to characterize dynamic processes to support our general qualitative findings is by no means trivial due to substantial spatial autocorrelations and on-board calibrations that bias observations.

%

\section{Conclusion  \label{sec:conclusion}}

Our mean-centered differencing procedure represents a significant improvement over all prior approaches to using the intensity of nighttime lights as a means to estimating the characteristics of macroeconomic phenomena. Our method efficiently filters away the effects of well documented (and replicated in this study) mean value biases in remote sensing that are caused by ground or on-board sensor conditions. As a result, our approach can provide reliable and interesting metrics across time and space that qualitatively agree with evidence from other data sources. While it is certainly possible to enhance the obtained metrics with quantitative interpolation across socioeconomic datasets to micro- and macro-levels using regressions or other statistical, geospatial methods, we stress the value and importance of our meso-level observations of the regional and time specific flow patterns presented in this paper which, moreover, have not been discussed or considered in the prior literature.

To date, no studies have looked at variance in global nighttime luminosity data across time as a proxy for convergence and divergence phenomena or, for that matter, any theorized dynamic economic processes. Previous work has merely noted correlations between nighttime lights and aggregate static accounts data (e.g., GDP). As noted above, unlike our approach, these prior studies focused on aggregate nighttime lights panels which are, themselves, deeply flawed. One likely reason for the relative scarcity of studies like this one is the lack of examples of big data analysis in economics that leverage interesting higher statistical moments and patterns that are computed and analyzed rigorously, with rare exceptions in \cite{yoshikawa2015deflation, butaru2015risk, burdick2014data, giesecke2016}. We argue that more research is needed that makes use of big (complex) data and takes seriously the theoretical underpinnings of the aggregate models currently in use. This will not only lead to improvements in current theory, but also shows promise for validation, rejection, and, even, introduction of novel hypotheses and theory.

Passive remote sensing of the earth's surface using satellites can allow researchers to obtain real-time proxies for economic activity that are spatially and temporally well-defined. As opposed to the somewhat fictitious aggregate data generated by society's tax optimized accounting, where investments in production and sales of products can be artificially separated in offshore holding companies with country of origin labels attached that do not reflect the actual economic transactions, seeing is believing. Moreover, while most survey-based data can also be misleading due to sample biases and hidden information, here we have a global dataset with complete coverage of nighttime photon spillage from human activity with, arguably, leading real-time indicators of economic activity that can be useful for a more precise characterization of on the ground economic development as well as early crisis warnings when investments do not realize growth in economic output as, otherwise, expected. 

We observe both support for the broad $\sigma$-convergence theory in the post-Cold war period as well as strong divergence dynamics immediately following the financial crisis and Great Recession with, moreover, observable patterns of growth that are correlated in time with governmental policies introduced in the wake of economic shocks. While it can be argued that the social insurance of government transfers provides benefits by limiting the extent and magnitude of Schumpeterian damage during and immediately following a crisis, we observe that there is both a direct cost of generally lower growth in more successful areas, as well as an indirect cost of higher accumulation of benefits in more urban areas leading to flows of labor and capital into already densely populated regions while leaving the hinterland relatively more impoverished.  These metrics could be improved significantly with even more up-to-date data. Particularly, with real-time (daily), high-resolution feeds of satellite data, the improvements in precision for global activity anomaly detection would enable a vast portfolio of early crisis warning signaling systems. Governments, foundations, philanthropy, and NGOs could intervene in real time as, for instance, communities slip into decay.

We have stood for centuries, squinting skyward, learning from looking. One of the most striking and exciting byproducts of our approach is the manner in which our visual representations of human economic activity are immediately suggestive, drawing the inquisitive researcher's eye to conspicuous structures, asterisms of light.

\begin{backmatter}

\section*{Competing interests}
The authors declare that they have no competing interests.

\section*{Author's contributions}
ED and VZ conceived and designed the research. VZ acquired and analyzed the data.  ED and VZ interpreted the results. All authors discussed, wrote, and approved the final version of the manuscript.

\section*{Acknowledgements}

This work was completed in part with resources provided by the University of Chicago Research Computing Center and a generous gift from Facebook.
\bibliographystyle{bmc-mathphys}
\bibliography{lightbib}


\begin{thebibliography}{38}
\ifx \bisbn   \undefined \def \bisbn  #1{ISBN #1}\fi
\ifx \binits  \undefined \def \binits#1{#1}\fi
\ifx \bauthor  \undefined \def \bauthor#1{#1}\fi
\ifx \batitle  \undefined \def \batitle#1{#1}\fi
\ifx \bjtitle  \undefined \def \bjtitle#1{#1}\fi
\ifx \bvolume  \undefined \def \bvolume#1{\textbf{#1}}\fi
\ifx \byear  \undefined \def \byear#1{#1}\fi
\ifx \bissue  \undefined \def \bissue#1{#1}\fi
\ifx \bfpage  \undefined \def \bfpage#1{#1}\fi
\ifx \blpage  \undefined \def \blpage #1{#1}\fi
\ifx \burl  \undefined \def \burl#1{\textsf{#1}}\fi
\ifx \doiurl  \undefined \def \doiurl#1{\textsf{#1}}\fi
\ifx \betal  \undefined \def \betal{\textit{et al.}}\fi
\ifx \binstitute  \undefined \def \binstitute#1{#1}\fi
\ifx \binstitutionaled  \undefined \def \binstitutionaled#1{#1}\fi
\ifx \bctitle  \undefined \def \bctitle#1{#1}\fi
\ifx \beditor  \undefined \def \beditor#1{#1}\fi
\ifx \bpublisher  \undefined \def \bpublisher#1{#1}\fi
\ifx \bbtitle  \undefined \def \bbtitle#1{#1}\fi
\ifx \bedition  \undefined \def \bedition#1{#1}\fi
\ifx \bseriesno  \undefined \def \bseriesno#1{#1}\fi
\ifx \blocation  \undefined \def \blocation#1{#1}\fi
\ifx \bsertitle  \undefined \def \bsertitle#1{#1}\fi
\ifx \bsnm \undefined \def \bsnm#1{#1}\fi
\ifx \bsuffix \undefined \def \bsuffix#1{#1}\fi
\ifx \bparticle \undefined \def \bparticle#1{#1}\fi
\ifx \barticle \undefined \def \barticle#1{#1}\fi
\ifx \bconfdate \undefined \def \bconfdate #1{#1}\fi
\ifx \botherref \undefined \def \botherref #1{#1}\fi
\ifx \url \undefined \def \url#1{\textsf{#1}}\fi
\ifx \bchapter \undefined \def \bchapter#1{#1}\fi
\ifx \bbook \undefined \def \bbook#1{#1}\fi
\ifx \bcomment \undefined \def \bcomment#1{#1}\fi
\ifx \oauthor \undefined \def \oauthor#1{#1}\fi
\ifx \citeauthoryear \undefined \def \citeauthoryear#1{#1}\fi
\ifx \endbibitem  \undefined \def \endbibitem {}\fi
\ifx \bconflocation  \undefined \def \bconflocation#1{#1}\fi
\ifx \arxivurl  \undefined \def \arxivurl#1{\textsf{#1}}\fi
\csname PreBibitemsHook\endcsname

\bibitem{evans2010philosophy}
\begin{barticle}
\bauthor{\bsnm{Evans}, \binits{J.}},
\bauthor{\bsnm{Rzhetsky}, \binits{A.}}:
\batitle{Philosophy of science: Machine science}.
\bjtitle{Science (New York, NY)}
\bvolume{329}(\bissue{5990}),
\bfpage{399}
(\byear{2010})
\end{barticle}
\endbibitem

\bibitem{hey2009fourth}
\begin{botherref}
\oauthor{\bsnm{Hey}, \binits{A.J.}},
\oauthor{\bsnm{Tansley}, \binits{S.}},
\oauthor{\bsnm{Tolle}, \binits{K.M.}}, et al.:
The fourth paradigm: data-intensive scientific discovery
\textbf{1}
(2009)
\end{botherref}
\endbibitem

\bibitem{aad2012observation}
\begin{barticle}
\bauthor{\bsnm{Aad}, \binits{G.}},
\bauthor{\bsnm{Abajyan}, \binits{T.}},
\bauthor{\bsnm{Abbott}, \binits{B.}},
\bauthor{\bsnm{Abdallah}, \binits{J.}},
\bauthor{\bsnm{Khalek}, \binits{S.A.}},
\bauthor{\bsnm{Abdelalim}, \binits{A.}},
\bauthor{\bsnm{Abdinov}, \binits{O.}},
\bauthor{\bsnm{Aben}, \binits{R.}},
\bauthor{\bsnm{Abi}, \binits{B.}},
\bauthor{\bsnm{Abolins}, \binits{M.}}, \betal:
\batitle{Observation of a new particle in the search for the {S}tandard {M}odel
  {H}iggs boson with the {A}{T}{L}{A}{S} detector at the {L}{H}{C}}.
\bjtitle{Physics Letters B}
\bvolume{716}(\bissue{1}),
\bfpage{1}--\blpage{29}
(\byear{2012})
\end{barticle}
\endbibitem

\bibitem{abbott2016observation}
\begin{barticle}
\bauthor{\bsnm{Abbott}, \binits{B.}},
\bauthor{\bsnm{Abbott}, \binits{R.}},
\bauthor{\bsnm{Abbott}, \binits{T.}},
\bauthor{\bsnm{Abernathy}, \binits{M.}},
\bauthor{\bsnm{Acernese}, \binits{F.}},
\bauthor{\bsnm{Ackley}, \binits{K.}},
\bauthor{\bsnm{Adams}, \binits{C.}},
\bauthor{\bsnm{Adams}, \binits{T.}},
\bauthor{\bsnm{Addesso}, \binits{P.}},
\bauthor{\bsnm{Adhikari}, \binits{R.}}, \betal:
\batitle{Observation of {G}ravitational {W}aves from a {B}inary {B}lack {H}ole
  {M}erger}.
\bjtitle{Physical Review Letters}
\bvolume{116}(\bissue{6}),
\bfpage{061102}
(\byear{2016})
\end{barticle}
\endbibitem

\bibitem{covert2004integrating}
\begin{barticle}
\bauthor{\bsnm{Covert}, \binits{M.W.}},
\bauthor{\bsnm{Knight}, \binits{E.M.}},
\bauthor{\bsnm{Reed}, \binits{J.L.}},
\bauthor{\bsnm{Herrgard}, \binits{M.J.}},
\bauthor{\bsnm{Palsson}, \binits{B.O.}}:
\batitle{Integrating high-throughput and computational data elucidates
  bacterial networks}.
\bjtitle{Nature}
\bvolume{429}(\bissue{6987}),
\bfpage{92}--\blpage{96}
(\byear{2004})
\end{barticle}
\endbibitem

\bibitem{conrad2014indirect}
\begin{botherref}
\oauthor{\bsnm{Conrad}, \binits{J.}}:
Indirect {D}etection of {W}{I}{M}{P} {D}ark {M}atter: a compact review.
arXiv preprint arXiv:1411.1925
(2014)
\end{botherref}
\endbibitem

\bibitem{beltran2009deducing}
\begin{barticle}
\bauthor{\bsnm{Beltran}, \binits{M.}},
\bauthor{\bsnm{Hooper}, \binits{D.}},
\bauthor{\bsnm{Kolb}, \binits{E.W.}},
\bauthor{\bsnm{Krusberg}, \binits{Z.A.}}:
\batitle{Deducing the nature of dark matter from direct and indirect detection
  experiments in the absence of collider signatures of new physics}.
\bjtitle{Physical Review D}
\bvolume{80}(\bissue{4}),
\bfpage{043509}
(\byear{2009})
\end{barticle}
\endbibitem

\bibitem{united2006current}
\begin{botherref}
\oauthor{\bsnm{BLS}}:
Current {P}opulation {S}urvey: Design and methodology.
U.S. Department of Labor
(2006)
\end{botherref}
\endbibitem

\bibitem{kaplan2012interstate}
\begin{barticle}
\bauthor{\bsnm{Kaplan}, \binits{G.}},
\bauthor{\bsnm{Schulhofer-Wohl}, \binits{S.}}:
\batitle{Interstate migration has fallen less than you think: {C}nsequences of
  hot deck imputation in the current population survey}.
\bjtitle{Demography}
\bvolume{49}(\bissue{3}),
\bfpage{1061}--\blpage{1074}
(\byear{2012})
\end{barticle}
\endbibitem

\bibitem{nordhaus2006geography}
\begin{barticle}
\bauthor{\bsnm{Nordhaus}, \binits{W.D.}}:
\batitle{Geography and macroeconomics: {N}ew data and new findings}.
\bjtitle{Proceedings of the National Academy of Sciences of the United States
  of America}
\bvolume{103}(\bissue{10}),
\bfpage{3510}--\blpage{3517}
(\byear{2006})
\end{barticle}
\endbibitem

\bibitem{sinton2001accuracy}
\begin{barticle}
\bauthor{\bsnm{Sinton}, \binits{J.E.}}:
\batitle{Accuracy and reliability of {C}hina's energy statistics}.
\bjtitle{China Economic Review}
\bvolume{12}(\bissue{4}),
\bfpage{373}--\blpage{383}
(\byear{2001})
\end{barticle}
\endbibitem

\bibitem{rogoff2010growth}
\begin{barticle}
\bauthor{\bsnm{Rogoff}, \binits{K.}},
\bauthor{\bsnm{Reinhart}, \binits{C.}}:
\batitle{Growth in a {T}ime of {D}ebt}.
\bjtitle{American Economic Review}
\bvolume{100}(\bissue{2}),
\bfpage{573}--\blpage{8}
(\byear{2010})
\end{barticle}
\endbibitem

\bibitem{herndon2014does}
\begin{barticle}
\bauthor{\bsnm{Herndon}, \binits{T.}},
\bauthor{\bsnm{Ash}, \binits{M.}},
\bauthor{\bsnm{Pollin}, \binits{R.}}:
\batitle{Does high public debt consistently stifle economic growth? {A}
  critique of {R}einhart and {R}ogoff}.
\bjtitle{Cambridge journal of economics}
\bvolume{38}(\bissue{2}),
\bfpage{257}--\blpage{279}
(\byear{2014})
\end{barticle}
\endbibitem

\bibitem{Leamer2007}
\begin{botherref}
\oauthor{\bsnm{Leamer}, \binits{E.E.}}:
Chapter 67: {L}inking the {T}heory with the {D}ata: {T}hat is the {C}ore
  {P}roblem of {I}nternational {E}conomics.
Handbook of Econometrics,
vol. 6, Part A,
pp. 4587--4606.
Elsevier
(2007).
doi:\doiurl{10.1016/S1573-4412(07)06067-9}
\end{botherref}
\endbibitem

\bibitem{ghosh2010shedding}
\begin{botherref}
\oauthor{\bsnm{Ghosh}, \binits{T.}},
\oauthor{\bsnm{L~Powell}, \binits{R.}},
\oauthor{\bsnm{D~Elvidge}, \binits{C.}},
\oauthor{\bsnm{E~Baugh}, \binits{K.}},
\oauthor{\bsnm{C~Sutton}, \binits{P.}},
\oauthor{\bsnm{Anderson}, \binits{S.}}:
Shedding light on the global distribution of economic activity.
The Open Geography Journal
\textbf{3}(1)
(2010)
\end{botherref}
\endbibitem

\bibitem{levin2012high}
\begin{barticle}
\bauthor{\bsnm{Levin}, \binits{N.}},
\bauthor{\bsnm{Duke}, \binits{Y.}}:
\batitle{High spatial resolution night-time light images for demographic and
  socio-economic studies}.
\bjtitle{Remote Sensing of Environment}
\bvolume{119},
\bfpage{1}--\blpage{10}
(\byear{2012})
\end{barticle}
\endbibitem

\bibitem{henderson2012measuring}
\begin{botherref}
\oauthor{\bsnm{Henderson}, \binits{J.V.}},
\oauthor{\bsnm{Storeygard}, \binits{A.}},
\oauthor{\bsnm{Weil}, \binits{D.N.}}:
Measuring {E}conomic {G}rowth from {O}uter {S}pace.
The American Economic Review,
994--1028
(2012)
\end{botherref}
\endbibitem

\bibitem{chen2011using}
\begin{barticle}
\bauthor{\bsnm{Chen}, \binits{X.}},
\bauthor{\bsnm{Nordhaus}, \binits{W.D.}}:
\batitle{Using luminosity data as a proxy for economic statistics}.
\bjtitle{Proceedings of the National Academy of Sciences}
\bvolume{108}(\bissue{21}),
\bfpage{8589}--\blpage{8594}
(\byear{2011})
\end{barticle}
\endbibitem

\bibitem{elvidge2012night}
\begin{barticle}
\bauthor{\bsnm{Elvidge}, \binits{C.D.}},
\bauthor{\bsnm{Baugh}, \binits{K.E.}},
\bauthor{\bsnm{Anderson}, \binits{S.}},
\bauthor{\bsnm{Sutton}, \binits{P.}},
\bauthor{\bsnm{Ghosh}, \binits{T.}}:
\batitle{The {N}ight {L}ight {D}evelopment {I}ndex ({N}{L}{D}{I}): a spatially
  explicit measure of human development from satellite data}.
\bjtitle{Social Geography}
\bvolume{7}(\bissue{1}),
\bfpage{23}--\blpage{35}
(\byear{2012})
\end{barticle}
\endbibitem

\bibitem{pinkovskiy2016lights}
\begin{barticle}
\bauthor{\bsnm{Pinkovskiy}, \binits{M.}},
\bauthor{\bsnm{Sala-i-Martin}, \binits{X.}}:
\batitle{Lights, {C}amera… {I}ncome! {I}lluminating the {N}ational
  {A}ccounts-{H}ousehold {S}urveys {D}ebate}.
\bjtitle{The Quarterly Journal of Economics}
\bvolume{131}(\bissue{2}),
\bfpage{579}--\blpage{631}
(\byear{2016})
\end{barticle}
\endbibitem

\bibitem{nordhaus2015sharper}
\begin{barticle}
\bauthor{\bsnm{Nordhaus}, \binits{W.}},
\bauthor{\bsnm{Chen}, \binits{X.}}:
\batitle{A sharper image? {E}stimates of the precision of nighttime lights as a
  proxy for economic statistics}.
\bjtitle{Journal of Economic Geography}
\bvolume{15}(\bissue{1}),
\bfpage{217}--\blpage{246}
(\byear{2015})
\end{barticle}
\endbibitem

\bibitem{friedman1992old}
\begin{barticle}
\bauthor{\bsnm{Friedman}, \binits{M.}}:
\batitle{Do old fallacies ever die?}
\bjtitle{Journal of Economics Literature}
\bvolume{30},
\bfpage{2129}--\blpage{2132}
(\byear{1992})
\end{barticle}
\endbibitem

\bibitem{young2008sigma}
\begin{barticle}
\bauthor{\bsnm{Young}, \binits{A.T.}},
\bauthor{\bsnm{Higgins}, \binits{M.J.}},
\bauthor{\bsnm{Levy}, \binits{D.}}:
\batitle{Sigma convergence versus beta convergence: {E}vidence from {U}{S}
  county-level data}.
\bjtitle{Journal of Money, Credit and Banking}
\bvolume{40}(\bissue{5}),
\bfpage{1083}--\blpage{1093}
(\byear{2008})
\end{barticle}
\endbibitem

\bibitem{piketty2014capital}
\begin{botherref}
\oauthor{\bsnm{Piketty}, \binits{T.}}:
Capital in the twenty-first century.
Cambridge, MA, London
(2014)
\end{botherref}
\endbibitem

\bibitem{durlauf1999new}
\begin{barticle}
\bauthor{\bsnm{Durlauf}, \binits{S.N.}},
\bauthor{\bsnm{Quah}, \binits{D.T.}}:
\batitle{The new empirics of economic growth}.
\bjtitle{Handbook of macroeconomics}
\bvolume{1},
\bfpage{235}--\blpage{308}
(\byear{1999})
\end{barticle}
\endbibitem

\bibitem{easterly}
\begin{barticle}
\bauthor{\bsnm{Easterly}, \binits{W.}},
\bauthor{\bsnm{Levine}, \binits{R.}}:
\batitle{It's {N}ot {F}actor {A}ccumulation: {S}tylized {F}acts and {G}rowth
  {M}odels}.
\bjtitle{The World Bank Economic Review}
\bvolume{15}(\bissue{2}),
\bfpage{177}--\blpage{219}
(\byear{2001})
\end{barticle}
\endbibitem

\bibitem{nordhaus1996real}
\begin{bchapter}
\bauthor{\bsnm{Nordhaus}, \binits{W.D.}}:
\bctitle{Do real-output and real-wage measures capture reality? {T}he history
  of lighting suggests not}.
In: \bbtitle{The Economics of New Goods},
pp. \bfpage{27}--\blpage{70}.
\bpublisher{University of Chicago Press},
\blocation{Chicago}
(\byear{1996})
\end{bchapter}
\endbibitem

\bibitem{rodrik2013unconditional}
\begin{botherref}
\oauthor{\bsnm{Rodrik}, \binits{D.}}:
Unconditional {C}onvergence in {M}anufacturing.
The Quarterly Journal of Economics,
1--40
(2013)
\end{botherref}
\endbibitem

\bibitem{sala1996classical}
\begin{botherref}
\oauthor{\bsnm{Sala-i-Martin}, \binits{X.X.}}:
The classical approach to convergence analysis.
The economic journal,
1019--1036
(1996)
\end{botherref}
\endbibitem

\bibitem{doll2008ciesin}
\begin{botherref}
\oauthor{\bsnm{Doll}, \binits{C.N.}}:
{C}{I}{E}{S}{I}{N} thematic guide to night-time light remote sensing and its
  applications.
Center for International Earth Science Information Network of Columbia
  University, Palisades, NY
(2008)
\end{botherref}
\endbibitem

\bibitem{baugh2010development}
\begin{barticle}
\bauthor{\bsnm{Baugh}, \binits{K.}},
\bauthor{\bsnm{Elvidge}, \binits{C.D.}},
\bauthor{\bsnm{Ghosh}, \binits{T.}},
\bauthor{\bsnm{Ziskin}, \binits{D.}}:
\batitle{Development of a 2009 stable lights product using
  {D}{M}{S}{P}-{O}{L}{S} data}.
\bjtitle{Proceedings of the Asia-Pacific Advanced Network}
\bvolume{30},
\bfpage{114}--\blpage{130}
(\byear{2010})
\end{barticle}
\endbibitem

\bibitem{free2014trends}
\begin{barticle}
\bauthor{\bsnm{Free}, \binits{M.}},
\bauthor{\bsnm{Sun}, \binits{B.}}:
\batitle{Trends in {U}{S} total cloud cover from a homogeneity-adjusted
  dataset}.
\bjtitle{Journal of Climate}
\bvolume{27}(\bissue{13}),
\bfpage{4959}--\blpage{4969}
(\byear{2014})
\end{barticle}
\endbibitem

\bibitem{norris2015empirical}
\begin{barticle}
\bauthor{\bsnm{Norris}, \binits{J.R.}},
\bauthor{\bsnm{Evan}, \binits{A.T.}}:
\batitle{Empirical removal of artifacts from the {I}{S}{C}{C}{P} and
  {P}{A}{T}{M}{O}{S}-x satellite cloud records}.
\bjtitle{Journal of Atmospheric and Oceanic Technology}
\bvolume{32}(\bissue{4}),
\bfpage{691}--\blpage{702}
(\byear{2015})
\end{barticle}
\endbibitem

\bibitem{lawrence2013global}
\begin{barticle}
\bauthor{\bsnm{Lawrence}, \binits{S.}},
\bauthor{\bsnm{Liu}, \binits{Q.}},
\bauthor{\bsnm{Yakovenko}, \binits{V.M.}}:
\batitle{Global inequality in energy consumption from 1980 to 2010}.
\bjtitle{Entropy}
\bvolume{15}(\bissue{12}),
\bfpage{5565}--\blpage{5579}
(\byear{2013})
\end{barticle}
\endbibitem

\bibitem{yoshikawa2015deflation}
\begin{botherref}
\oauthor{\bsnm{Yoshikawa}, \binits{H.}},
\oauthor{\bsnm{Aoyama}, \binits{H.}},
\oauthor{\bsnm{Fujiwara}, \binits{Y.}},
\oauthor{\bsnm{Iyetomi}, \binits{H.}}:
Deflation/{I}nflation {D}ynamics: {A}nalysis based on micro prices.
Available at SSRN 2565599
(2015)
\end{botherref}
\endbibitem

\bibitem{butaru2015risk}
\begin{botherref}
\oauthor{\bsnm{Butaru}, \binits{F.}},
\oauthor{\bsnm{Chen}, \binits{Q.}},
\oauthor{\bsnm{Clark}, \binits{B.}},
\oauthor{\bsnm{Das}, \binits{S.}},
\oauthor{\bsnm{Lo}, \binits{A.W.}},
\oauthor{\bsnm{Siddique}, \binits{A.}}:
Risk and {R}isk {M}anagement in the {C}redit {C}ard {I}ndustry.
Technical report,
National Bureau of Economic Research
(2015)
\end{botherref}
\endbibitem

\bibitem{burdick2014data}
\begin{bchapter}
\bauthor{\bsnm{Burdick}, \binits{D.}},
\bauthor{\bsnm{Franklin}, \binits{M.}},
\bauthor{\bsnm{Issler}, \binits{P.}},
\bauthor{\bsnm{Krishnamurthy}, \binits{R.}},
\bauthor{\bsnm{Popa}, \binits{L.}},
\bauthor{\bsnm{Raschid}, \binits{L.}},
\bauthor{\bsnm{Stanton}, \binits{R.}},
\bauthor{\bsnm{Wallace}, \binits{N.}}:
\bctitle{Data {S}cience {C}hallenges in {R}eal {E}state {A}sset and {C}apital
  {M}arkets}.
In: \bbtitle{Proceedings of the International Workshop on Data Science for
  Macro-Modeling},
pp. \bfpage{1}--\blpage{5}
(\byear{2014}).
\bcomment{ACM}
\end{bchapter}
\endbibitem

\bibitem{giesecke2016}
\begin{botherref}
\oauthor{\bsnm{Giesecke}, \binits{K.}},
\oauthor{\bsnm{Sirignano}, \binits{J.}},
\oauthor{\bsnm{Sadhwani}, \binits{A.}}:
Deep {L}earning for {M}ortgage {R}isk.
Working Paper, Stanford University
(2016)
\end{botherref}
\endbibitem

\end{thebibliography}

\newcommand{\BMCxmlcomment}[1]{}

\BMCxmlcomment{

<refgrp>

<bibl id="B1">
  <title><p>Philosophy of science: Machine science</p></title>
  <aug>
    <au><snm>Evans</snm><fnm>J</fnm></au>
    <au><snm>Rzhetsky</snm><fnm>A</fnm></au>
  </aug>
  <source>Science (New York, NY)</source>
  <publisher>NIH Public Access</publisher>
  <pubdate>2010</pubdate>
  <volume>329</volume>
  <issue>5990</issue>
  <fpage>399</fpage>
</bibl>

<bibl id="B2">
  <title><p>The fourth paradigm: data-intensive scientific
  discovery</p></title>
  <aug>
    <au><snm>Hey</snm><fnm>AJ</fnm></au>
    <au><snm>Tansley</snm><fnm>S</fnm></au>
    <au><snm>Tolle</snm><fnm>KM</fnm></au>
    <au><cnm>others</cnm></au>
  </aug>
  <publisher>Microsoft research Redmond, WA</publisher>
  <pubdate>2009</pubdate>
  <volume>1</volume>
</bibl>

<bibl id="B3">
  <title><p>Observation of a new particle in the search for the {S}tandard
  {M}odel {H}iggs boson with the {A}{T}{L}{A}{S} detector at the
  {L}{H}{C}</p></title>
  <aug>
    <au><snm>Aad</snm><fnm>G</fnm></au>
    <au><snm>Abajyan</snm><fnm>T</fnm></au>
    <au><snm>Abbott</snm><fnm>B</fnm></au>
    <au><snm>Abdallah</snm><fnm>J</fnm></au>
    <au><snm>Khalek</snm><fnm>SA</fnm></au>
    <au><snm>Abdelalim</snm><fnm>AA</fnm></au>
    <au><snm>Abdinov</snm><fnm>O</fnm></au>
    <au><snm>Aben</snm><fnm>R</fnm></au>
    <au><snm>Abi</snm><fnm>B</fnm></au>
    <au><snm>Abolins</snm><fnm>M</fnm></au>
    <au><cnm>others</cnm></au>
  </aug>
  <source>Physics Letters B</source>
  <publisher>Elsevier</publisher>
  <pubdate>2012</pubdate>
  <volume>716</volume>
  <issue>1</issue>
  <fpage>1</fpage>
  <lpage>-29</lpage>
</bibl>

<bibl id="B4">
  <title><p>Observation of {G}ravitational {W}aves from a {B}inary {B}lack
  {H}ole {M}erger</p></title>
  <aug>
    <au><snm>Abbott</snm><fnm>BP</fnm></au>
    <au><snm>Abbott</snm><fnm>R</fnm></au>
    <au><snm>Abbott</snm><fnm>TD</fnm></au>
    <au><snm>Abernathy</snm><fnm>MR</fnm></au>
    <au><snm>Acernese</snm><fnm>F</fnm></au>
    <au><snm>Ackley</snm><fnm>K</fnm></au>
    <au><snm>Adams</snm><fnm>C</fnm></au>
    <au><snm>Adams</snm><fnm>T</fnm></au>
    <au><snm>Addesso</snm><fnm>P</fnm></au>
    <au><snm>Adhikari</snm><fnm>RX</fnm></au>
    <au><cnm>others</cnm></au>
  </aug>
  <source>Physical Review Letters</source>
  <publisher>APS</publisher>
  <pubdate>2016</pubdate>
  <volume>116</volume>
  <issue>6</issue>
  <fpage>061102</fpage>
</bibl>

<bibl id="B5">
  <title><p>Integrating high-throughput and computational data elucidates
  bacterial networks</p></title>
  <aug>
    <au><snm>Covert</snm><fnm>MW</fnm></au>
    <au><snm>Knight</snm><fnm>EM</fnm></au>
    <au><snm>Reed</snm><fnm>JL</fnm></au>
    <au><snm>Herrgard</snm><fnm>MJ</fnm></au>
    <au><snm>Palsson</snm><fnm>BO</fnm></au>
  </aug>
  <source>Nature</source>
  <publisher>Nature Publishing Group</publisher>
  <pubdate>2004</pubdate>
  <volume>429</volume>
  <issue>6987</issue>
  <fpage>92</fpage>
  <lpage>-96</lpage>
</bibl>

<bibl id="B6">
  <title><p>Indirect {D}etection of {W}{I}{M}{P} {D}ark {M}atter: a compact
  review</p></title>
  <aug>
    <au><snm>Conrad</snm><fnm>J</fnm></au>
  </aug>
  <source>arXiv preprint arXiv:1411.1925</source>
  <pubdate>2014</pubdate>
</bibl>

<bibl id="B7">
  <title><p>Deducing the nature of dark matter from direct and indirect
  detection experiments in the absence of collider signatures of new
  physics</p></title>
  <aug>
    <au><snm>Beltran</snm><fnm>M</fnm></au>
    <au><snm>Hooper</snm><fnm>D</fnm></au>
    <au><snm>Kolb</snm><fnm>EW</fnm></au>
    <au><snm>Krusberg</snm><fnm>ZA</fnm></au>
  </aug>
  <source>Physical Review D</source>
  <publisher>APS</publisher>
  <pubdate>2009</pubdate>
  <volume>80</volume>
  <issue>4</issue>
  <fpage>043509</fpage>
</bibl>

<bibl id="B8">
  <title><p>Current {P}opulation {S}urvey: Design and methodology</p></title>
  <aug>
    <au><cnm>BLS</cnm></au>
  </aug>
  <publisher>U.S. Department of Labor</publisher>
  <series><title><p>Technical paper</p></title></series>
  <pubdate>2006</pubdate>
</bibl>

<bibl id="B9">
  <title><p>Interstate migration has fallen less than you think: {C}nsequences
  of hot deck imputation in the Current Population Survey</p></title>
  <aug>
    <au><snm>Kaplan</snm><fnm>G</fnm></au>
    <au><snm>Schulhofer Wohl</snm><fnm>S</fnm></au>
  </aug>
  <source>Demography</source>
  <publisher>Springer</publisher>
  <pubdate>2012</pubdate>
  <volume>49</volume>
  <issue>3</issue>
  <fpage>1061</fpage>
  <lpage>-1074</lpage>
</bibl>

<bibl id="B10">
  <title><p>Geography and macroeconomics: {N}ew data and new
  findings</p></title>
  <aug>
    <au><snm>Nordhaus</snm><fnm>WD</fnm></au>
  </aug>
  <source>Proceedings of the National Academy of Sciences of the United States
  of America</source>
  <publisher>National Acad Sciences</publisher>
  <pubdate>2006</pubdate>
  <volume>103</volume>
  <issue>10</issue>
  <fpage>3510</fpage>
  <lpage>-3517</lpage>
</bibl>

<bibl id="B11">
  <title><p>Accuracy and reliability of {C}hina's energy statistics</p></title>
  <aug>
    <au><snm>Sinton</snm><fnm>JE</fnm></au>
  </aug>
  <source>China Economic Review</source>
  <publisher>Elsevier</publisher>
  <pubdate>2001</pubdate>
  <volume>12</volume>
  <issue>4</issue>
  <fpage>373</fpage>
  <lpage>-383</lpage>
</bibl>

<bibl id="B12">
  <title><p>Growth in a {T}ime of {D}ebt</p></title>
  <aug>
    <au><snm>Rogoff</snm><fnm>K</fnm></au>
    <au><snm>Reinhart</snm><fnm>C</fnm></au>
  </aug>
  <source>American Economic Review</source>
  <pubdate>2010</pubdate>
  <volume>100</volume>
  <issue>2</issue>
  <fpage>573</fpage>
  <lpage>-8</lpage>
</bibl>

<bibl id="B13">
  <title><p>Does high public debt consistently stifle economic growth? {A}
  critique of {R}einhart and {R}ogoff</p></title>
  <aug>
    <au><snm>Herndon</snm><fnm>T</fnm></au>
    <au><snm>Ash</snm><fnm>M</fnm></au>
    <au><snm>Pollin</snm><fnm>R</fnm></au>
  </aug>
  <source>Cambridge journal of economics</source>
  <publisher>CPES</publisher>
  <pubdate>2014</pubdate>
  <volume>38</volume>
  <issue>2</issue>
  <fpage>257</fpage>
  <lpage>-279</lpage>
</bibl>

<bibl id="B14">
  <title><p>Chapter 67: {L}inking the {T}heory with the {D}ata: {T}hat is the
  {C}ore {P}roblem of {I}nternational {E}conomics</p></title>
  <aug>
    <au><snm>Leamer</snm><fnm>EE</fnm></au>
  </aug>
  <publisher>Elsevier</publisher>
  <editor>James J. Heckman and Edward E. Leamer</editor>
  <series><title><p>Handbook of Econometrics</p></title></series>
  <pubdate>2007</pubdate>
  <volume>6, Part A</volume>
  <fpage>4587</fpage>
  <lpage>4606</lpage>
</bibl>

<bibl id="B15">
  <title><p>Shedding light on the global distribution of economic
  activity</p></title>
  <aug>
    <au><snm>Ghosh</snm><fnm>T</fnm></au>
    <au><snm>L Powell</snm><fnm>R</fnm></au>
    <au><snm>D Elvidge</snm><fnm>C</fnm></au>
    <au><snm>E Baugh</snm><fnm>K</fnm></au>
    <au><snm>C Sutton</snm><fnm>P</fnm></au>
    <au><snm>Anderson</snm><fnm>S</fnm></au>
  </aug>
  <source>The Open Geography Journal</source>
  <pubdate>2010</pubdate>
  <volume>3</volume>
  <issue>1</issue>
</bibl>

<bibl id="B16">
  <title><p>High spatial resolution night-time light images for demographic and
  socio-economic studies</p></title>
  <aug>
    <au><snm>Levin</snm><fnm>N</fnm></au>
    <au><snm>Duke</snm><fnm>Y</fnm></au>
  </aug>
  <source>Remote Sensing of Environment</source>
  <publisher>Elsevier</publisher>
  <pubdate>2012</pubdate>
  <volume>119</volume>
  <fpage>1</fpage>
  <lpage>-10</lpage>
</bibl>

<bibl id="B17">
  <title><p>Measuring {E}conomic {G}rowth from {O}uter {S}pace</p></title>
  <aug>
    <au><snm>Henderson</snm><fnm>JV</fnm></au>
    <au><snm>Storeygard</snm><fnm>A</fnm></au>
    <au><snm>Weil</snm><fnm>DN</fnm></au>
  </aug>
  <source>The American Economic Review</source>
  <publisher>JSTOR</publisher>
  <pubdate>2012</pubdate>
  <fpage>994</fpage>
  <lpage>-1028</lpage>
</bibl>

<bibl id="B18">
  <title><p>Using luminosity data as a proxy for economic
  statistics</p></title>
  <aug>
    <au><snm>Chen</snm><fnm>X</fnm></au>
    <au><snm>Nordhaus</snm><fnm>WD</fnm></au>
  </aug>
  <source>Proceedings of the National Academy of Sciences</source>
  <publisher>National Acad Sciences</publisher>
  <pubdate>2011</pubdate>
  <volume>108</volume>
  <issue>21</issue>
  <fpage>8589</fpage>
  <lpage>-8594</lpage>
</bibl>

<bibl id="B19">
  <title><p>The {N}ight {L}ight {D}evelopment {I}ndex ({N}{L}{D}{I}): a
  spatially explicit measure of human development from satellite
  data</p></title>
  <aug>
    <au><snm>Elvidge</snm><fnm>CD</fnm></au>
    <au><snm>Baugh</snm><fnm>KE</fnm></au>
    <au><snm>Anderson</snm><fnm>SJ</fnm></au>
    <au><snm>Sutton</snm><fnm>PC</fnm></au>
    <au><snm>Ghosh</snm><fnm>T</fnm></au>
  </aug>
  <source>Social Geography</source>
  <publisher>Copernicus GmbH</publisher>
  <pubdate>2012</pubdate>
  <volume>7</volume>
  <issue>1</issue>
  <fpage>23</fpage>
  <lpage>-35</lpage>
</bibl>

<bibl id="B20">
  <title><p>Lights, {C}amera… {I}ncome! {I}lluminating the {N}ational
  {A}ccounts-{H}ousehold {S}urveys {D}ebate</p></title>
  <aug>
    <au><snm>Pinkovskiy</snm><fnm>M</fnm></au>
    <au><snm>Martin</snm><fnm>X</fnm></au>
  </aug>
  <source>The Quarterly Journal of Economics</source>
  <publisher>Oxford University Press</publisher>
  <pubdate>2016</pubdate>
  <volume>131</volume>
  <issue>2</issue>
  <fpage>579</fpage>
  <lpage>-631</lpage>
</bibl>

<bibl id="B21">
  <title><p>A sharper image? {E}stimates of the precision of nighttime lights
  as a proxy for economic statistics</p></title>
  <aug>
    <au><snm>Nordhaus</snm><fnm>W</fnm></au>
    <au><snm>Chen</snm><fnm>X</fnm></au>
  </aug>
  <source>Journal of Economic Geography</source>
  <publisher>Oxford Univ Press</publisher>
  <pubdate>2015</pubdate>
  <volume>15</volume>
  <issue>1</issue>
  <fpage>217</fpage>
  <lpage>-246</lpage>
</bibl>

<bibl id="B22">
  <title><p>Do old fallacies ever die?</p></title>
  <aug>
    <au><snm>Friedman</snm><fnm>M</fnm></au>
  </aug>
  <source>Journal of Economics Literature</source>
  <pubdate>1992</pubdate>
  <volume>30</volume>
  <fpage>2129</fpage>
  <lpage>-2132</lpage>
</bibl>

<bibl id="B23">
  <title><p>Sigma convergence versus beta convergence: {E}vidence from {U}{S}
  county-level data</p></title>
  <aug>
    <au><snm>Young</snm><fnm>AT</fnm></au>
    <au><snm>Higgins</snm><fnm>MJ</fnm></au>
    <au><snm>Levy</snm><fnm>D</fnm></au>
  </aug>
  <source>Journal of Money, Credit and Banking</source>
  <publisher>Wiley Online Library</publisher>
  <pubdate>2008</pubdate>
  <volume>40</volume>
  <issue>5</issue>
  <fpage>1083</fpage>
  <lpage>-1093</lpage>
</bibl>

<bibl id="B24">
  <title><p>Capital in the twenty-first century</p></title>
  <aug>
    <au><snm>Piketty</snm><fnm>T</fnm></au>
  </aug>
  <source>Cambridge, MA, London</source>
  <pubdate>2014</pubdate>
</bibl>

<bibl id="B25">
  <title><p>The new empirics of economic growth</p></title>
  <aug>
    <au><snm>Durlauf</snm><fnm>SN</fnm></au>
    <au><snm>Quah</snm><fnm>DT</fnm></au>
  </aug>
  <source>Handbook of macroeconomics</source>
  <publisher>Elsevier</publisher>
  <pubdate>1999</pubdate>
  <volume>1</volume>
  <fpage>235</fpage>
  <lpage>-308</lpage>
</bibl>

<bibl id="B26">
  <title><p>It's {N}ot {F}actor {A}ccumulation: {S}tylized {F}acts and {G}rowth
  {M}odels</p></title>
  <aug>
    <au><snm>Easterly</snm><fnm>W</fnm></au>
    <au><snm>Levine</snm><fnm>R</fnm></au>
  </aug>
  <source>The World Bank Economic Review</source>
  <publisher>Oxford University Press</publisher>
  <pubdate>2001</pubdate>
  <volume>15</volume>
  <issue>2</issue>
  <fpage>177</fpage>
  <lpage>219</lpage>
</bibl>

<bibl id="B27">
  <title><p>Do real-output and real-wage measures capture reality? {T}he
  history of lighting suggests not</p></title>
  <aug>
    <au><snm>Nordhaus</snm><fnm>WD</fnm></au>
  </aug>
  <source>The economics of new goods</source>
  <publisher>Chicago: University of Chicago Press</publisher>
  <pubdate>1996</pubdate>
  <fpage>27</fpage>
  <lpage>-70</lpage>
</bibl>

<bibl id="B28">
  <title><p>Unconditional {C}onvergence in {M}anufacturing</p></title>
  <aug>
    <au><snm>Rodrik</snm><fnm>D</fnm></au>
  </aug>
  <source>The Quarterly Journal of Economics</source>
  <publisher>Oxford University Press</publisher>
  <pubdate>2013</pubdate>
  <fpage>1</fpage>
  <lpage>-40</lpage>
</bibl>

<bibl id="B29">
  <title><p>The classical approach to convergence analysis</p></title>
  <aug>
    <au><snm>Martin</snm><fnm>XX</fnm></au>
  </aug>
  <source>The economic journal</source>
  <publisher>JSTOR</publisher>
  <pubdate>1996</pubdate>
  <fpage>1019</fpage>
  <lpage>-1036</lpage>
</bibl>

<bibl id="B30">
  <title><p>{C}{I}{E}{S}{I}{N} thematic guide to night-time light remote
  sensing and its applications</p></title>
  <aug>
    <au><snm>Doll</snm><fnm>CN</fnm></au>
  </aug>
  <source>Center for International Earth Science Information Network of
  Columbia University, Palisades, NY</source>
  <pubdate>2008</pubdate>
</bibl>

<bibl id="B31">
  <title><p>Development of a 2009 stable lights product using
  {D}{M}{S}{P}-{O}{L}{S} data</p></title>
  <aug>
    <au><snm>Baugh</snm><fnm>K</fnm></au>
    <au><snm>Elvidge</snm><fnm>CD</fnm></au>
    <au><snm>Ghosh</snm><fnm>T</fnm></au>
    <au><snm>Ziskin</snm><fnm>D</fnm></au>
  </aug>
  <source>Proceedings of the Asia-Pacific Advanced Network</source>
  <pubdate>2010</pubdate>
  <volume>30</volume>
  <fpage>114</fpage>
  <lpage>-130</lpage>
</bibl>

<bibl id="B32">
  <title><p>Trends in {U}{S} total cloud cover from a homogeneity-adjusted
  dataset</p></title>
  <aug>
    <au><snm>Free</snm><fnm>M</fnm></au>
    <au><snm>Sun</snm><fnm>B</fnm></au>
  </aug>
  <source>Journal of Climate</source>
  <pubdate>2014</pubdate>
  <volume>27</volume>
  <issue>13</issue>
  <fpage>4959</fpage>
  <lpage>-4969</lpage>
</bibl>

<bibl id="B33">
  <title><p>Empirical removal of artifacts from the {I}{S}{C}{C}{P} and
  {P}{A}{T}{M}{O}{S}-x satellite cloud records</p></title>
  <aug>
    <au><snm>Norris</snm><fnm>JR</fnm></au>
    <au><snm>Evan</snm><fnm>AT</fnm></au>
  </aug>
  <source>Journal of Atmospheric and Oceanic Technology</source>
  <pubdate>2015</pubdate>
  <volume>32</volume>
  <issue>4</issue>
  <fpage>691</fpage>
  <lpage>-702</lpage>
</bibl>

<bibl id="B34">
  <title><p>Global inequality in energy consumption from 1980 to
  2010</p></title>
  <aug>
    <au><snm>Lawrence</snm><fnm>S</fnm></au>
    <au><snm>Liu</snm><fnm>Q</fnm></au>
    <au><snm>Yakovenko</snm><fnm>VM</fnm></au>
  </aug>
  <source>Entropy</source>
  <publisher>Multidisciplinary Digital Publishing Institute</publisher>
  <pubdate>2013</pubdate>
  <volume>15</volume>
  <issue>12</issue>
  <fpage>5565</fpage>
  <lpage>-5579</lpage>
</bibl>

<bibl id="B35">
  <title><p>Deflation/{I}nflation {D}ynamics: {A}nalysis based on micro
  prices</p></title>
  <aug>
    <au><snm>Yoshikawa</snm><fnm>H</fnm></au>
    <au><snm>Aoyama</snm><fnm>H</fnm></au>
    <au><snm>Fujiwara</snm><fnm>Y</fnm></au>
    <au><snm>Iyetomi</snm><fnm>H</fnm></au>
  </aug>
  <source>Available at SSRN 2565599</source>
  <pubdate>2015</pubdate>
</bibl>

<bibl id="B36">
  <title><p>Risk and {R}isk {M}anagement in the {C}redit {C}ard
  {I}ndustry</p></title>
  <aug>
    <au><snm>Butaru</snm><fnm>F</fnm></au>
    <au><snm>Chen</snm><fnm>Q</fnm></au>
    <au><snm>Clark</snm><fnm>B</fnm></au>
    <au><snm>Das</snm><fnm>S</fnm></au>
    <au><snm>Lo</snm><fnm>AW</fnm></au>
    <au><snm>Siddique</snm><fnm>A</fnm></au>
  </aug>
  <pubdate>2015</pubdate>
</bibl>

<bibl id="B37">
  <title><p>Data {S}cience {C}hallenges in {R}eal {E}state {A}sset and
  {C}apital {M}arkets</p></title>
  <aug>
    <au><snm>Burdick</snm><fnm>D</fnm></au>
    <au><snm>Franklin</snm><fnm>M</fnm></au>
    <au><snm>Issler</snm><fnm>P</fnm></au>
    <au><snm>Krishnamurthy</snm><fnm>R</fnm></au>
    <au><snm>Popa</snm><fnm>L</fnm></au>
    <au><snm>Raschid</snm><fnm>L</fnm></au>
    <au><snm>Stanton</snm><fnm>R</fnm></au>
    <au><snm>Wallace</snm><fnm>N</fnm></au>
  </aug>
  <source>Proceedings of the International Workshop on Data Science for
  Macro-Modeling</source>
  <pubdate>2014</pubdate>
  <fpage>1</fpage>
  <lpage>-5</lpage>
</bibl>

<bibl id="B38">
  <title><p>Deep {L}earning for {M}ortgage {R}isk</p></title>
  <aug>
    <au><snm>Giesecke</snm><fnm>K</fnm></au>
    <au><snm>Sirignano</snm><fnm>J</fnm></au>
    <au><snm>Sadhwani</snm><fnm>A</fnm></au>
  </aug>
  <source>Working Paper, Stanford University</source>
  <pubdate>2016</pubdate>
</bibl>

</refgrp>
} 



\end{backmatter}

\end{document}